\documentclass[prl,twocolumn,nopacs,amsmath,amssymb,superscriptaddress]{revtex4-2}

\usepackage{amsmath}
\usepackage{amssymb}
\usepackage{amsbsy}
\usepackage{graphicx}
\usepackage{xcolor}

\usepackage{mathrsfs}

\usepackage{enumerate}
\usepackage{mathtools}

\usepackage{dcolumn}

\newcommand{\moy}[1]{\left\langle #1 \right\rangle}
\newcommand{\dd}[0]{\mathrm{d}}

\newcommand{\erf}[0]{\text{erf}}

\usepackage{soul}

\newcommand{\dt}[2]{\ensuremath{\frac{\dd #1}{\dd #2}}}

\def\e{e}

\newcommand{\dep}[2]{\ensuremath{\frac{\partial #1}{\partial #2}}}

\newcommand{\sg}[1]{\ensuremath{\mathrm{sign}\left( #1 \right)}}
\newcommand{\abs}[1]{\ensuremath{\left| #1 \right|}}

\DeclareMathOperator{\erfc}{erfc}

\definecolor{darkblue}{rgb}{0,0,0.6}
\definecolor{darkred}{rgb}{0.6,0,0}

\usepackage[colorlinks=true,urlcolor=darkblue,citecolor=darkblue,linkcolor=darkred]{hyperref}

\def\Dt{\tilde{D}}

\def\sprb{\sigma'(\rho)}

\def\rb{\bar\rho}

\def\QT{Q_T}
\def\Qt{Q_t}

\def\rb{\bar\rho}

\def\sprb{\sigma'(\rho)}

\def\QT{Q_T}
\def\Qt{Q_t}

\def\D{\mathcal{D}}

\def\Fk{\mathcal{G}}

\def\rhom{\rho_{\mathrm{micro}}}
\def\jm{j_{\mathrm{micro}}}

\newcommand{\eqlaw}{\ensuremath{\overset{(\mathrm{law})}{=}}}

\begin{document}

\title{Tracer and current fluctuations in driven diffusive systems}

\author{Th\'eotim Berlioz}
\affiliation{Sorbonne Universit\'e, CNRS, Laboratoire de Physique Th\'eorique de la Mati\`ere Condens\'ee (LPTMC), 4 Place Jussieu, 75005 Paris, France}

\author{Olivier B\'enichou}
\affiliation{Sorbonne Universit\'e, CNRS, Laboratoire de Physique Th\'eorique de la Mati\`ere Condens\'ee (LPTMC), 4 Place Jussieu, 75005 Paris, France}

\author{Aur\'elien Grabsch}
\affiliation{Sorbonne Universit\'e, CNRS, Laboratoire de Physique Th\'eorique de la Mati\`ere Condens\'ee (LPTMC), 4 Place Jussieu, 75005 Paris, France}

\begin{abstract}
    Interacting particles diffusing in single-file is a fundamental model of transport in narrow channels where particles cannot bypass each other. An important result has been obtained by Kollmann [Phys. Rev. Lett. \textbf{90}, 180602 (2003)] for the mean square displacement of a tracer for any single-file model. It applies to any diffusive system, in particular the notable classes of colloidal systems and diffusive stochastic lattice gases.
    Since then, no analog result has been obtained in the important case where the particles are driven by an external field. Here, we fill this gap and determine the fluctuations and the skewness of the tracer's position for any driven diffusive system.
    In addition, we also consider a variety of important observables such as the integrated current, the response of the system to the perturbation induced by the displacement of the tracer, and the correlations between several tracers. Furthermore, we also unveil fundamental relations underlying the out-of-equilibrium dynamics of driven diffusive systems.
    This work constitutes a step toward the determination of the full distribution of all these observables in driven one-dimensional systems of interacting particles.
\end{abstract}

\maketitle

\let\oldaddcontentsline\addcontentsline
\renewcommand{\addcontentsline}[3]{}

\emph{Introduction.---} The investigation of the dynamical properties of interacting particle systems in both equilibrium and non-equilibrium settings has been a prominent area of research in the last decades~\cite{Spohn:1991,Evans:2005,Derrida:2007,Chou:2011,Bertini:2015}. A paradigmatic example is the single-file model, which describes particles diffusing in narrow channels, so that they cannot bypass each over. 
In this context, two quantities have attracted a lot of attention: the position $X_t$ of a tracer, and the integrated current (which measures the total particles flux through the origin).
It was shown that at long times the mean squared displacement of a tracer particle grows sublinearly with time, $\moy{X_T^2} \propto T^{1/2}$~\cite{Harris:1965,Levitt:1973,Arratia:1983}. This is a general property, which has been observed experimentally in various single-file systems~\cite{Perkins:1994,Hahn:1996,Wei:2000,Lin:2005}. However the precise determination of the prefactor, which contains all the dependence on the physical properties of the system, such as the nature of the interaction, the value of the density, or the initial condition, is subtle.

For a general single-file system, starting from an equilibrium distribution with mean density $\rb$, the mean-squared displacement (MSD) of the tracer, including the prefactor, has first been determined for reflecting Brownian particles~\cite{Harris:1965}~\footnote{The model of reflecting Brownian particles refers to the 1D system of Brownian particles with a hardcore exclusion constraint.}, and later for the Simple Exclusion Process (SEP)~\cite{Arratia:1983}~\footnote{The SEP is a models of particles on a lattice which randomly hop with symmetric rates to neighbouring sites, with a hardcore exclusion constraint.}, which is a paradigmatic model of single-file diffusion. A major progress has been accomplished by Kollman~\cite{Kollmann:2003} who obtained the result for \textit{any} single-file system~\footnote{The mean squared displacement of a tracer in any single-file system was first derived by Kollmann in~\cite{Kollmann:2003}. This result was rewritten in terms of $D(\rho)$ and $\sigma(\rho)$ in~\cite{Krapivsky:2014}. This is the expression that we give in Eq.~\eqref{eq:Kappa2Kollmann}.},
\begin{equation}
    \label{eq:Kappa2Kollmann}
    \moy{X_T^2} \underset{T \to \infty}{\simeq}
    \frac{\sigma(\rb)}{\rb^2 \sqrt{\pi D(\rb)}} \sqrt{T}
    \:,
\end{equation}
where $D(\rho)$  is the collective diffusion coefficient, and $\sigma(\rho)$ the mobility, which controls both the current fluctuations, and the response of the current to a small drive.
We stress that this result applies not only to colloidal systems, but to any diffusive system (i.e. for which the density $\rho$ and the current $j$ obey Fick's law $j = - D(\rho) \partial_x \rho$) including for instance stochastic lattice gases. Equation~\eqref{eq:Kappa2Kollmann} has been very influential in the field of single-file diffusion, both at the theoretical~\cite{Felderhof:2009,Barkai:2009,Lizana:2010,Krapivsky:2014,Krapivsky:2015a,Hartich:2021,Dolai:2020,Akintunde:2024} and experimental level~\cite{Lutz:2004,Lutz:2004b,Lin:2005,Coste:2010}.

The fluctuations of the integrated current $\QT$ through the origin, defined as the number of particles that have crossed the origin up to time $T$ from left to right minus those from right to left, have also been determined, and take a similar form~\cite{Krapivsky:2012},
\begin{equation}
    \label{eq:Kappa2Qt}
    \moy{\QT^2} \underset{T \to \infty}{\simeq}
    \frac{\sigma(\rb)}{\sqrt{\pi D(\rb)}} \sqrt{T}
    \:.
\end{equation}

The distributions  of both $X_T$ and $Q_T$ have been shown to be non-Gaussian~\cite{Derrida:2009a,Krapivsky:2015a}. Recently, there has been a lot of interest in characterizing these distributions beyond the variance, particularly through higher-order cumulants, which quantify deviations from Gaussianity.
All the cumulants have first been determined for specific models, such as reflecting Brownian particles~\cite{Hegde:2014,Krapivsky:2014,Krapivsky:2015a,Sadhu:2015}, and the SEP~\cite{Derrida:2009,Derrida:2009a,Illien:2013,Imamura:2017,Imamura:2021}.
For general single-file systems the question is more complex and, beyond the MSD, only the fourth cumulant has been determined recently~\cite{Grabsch:2024b}.

\begin{figure}
    \centering
    \includegraphics[width=0.8\columnwidth]{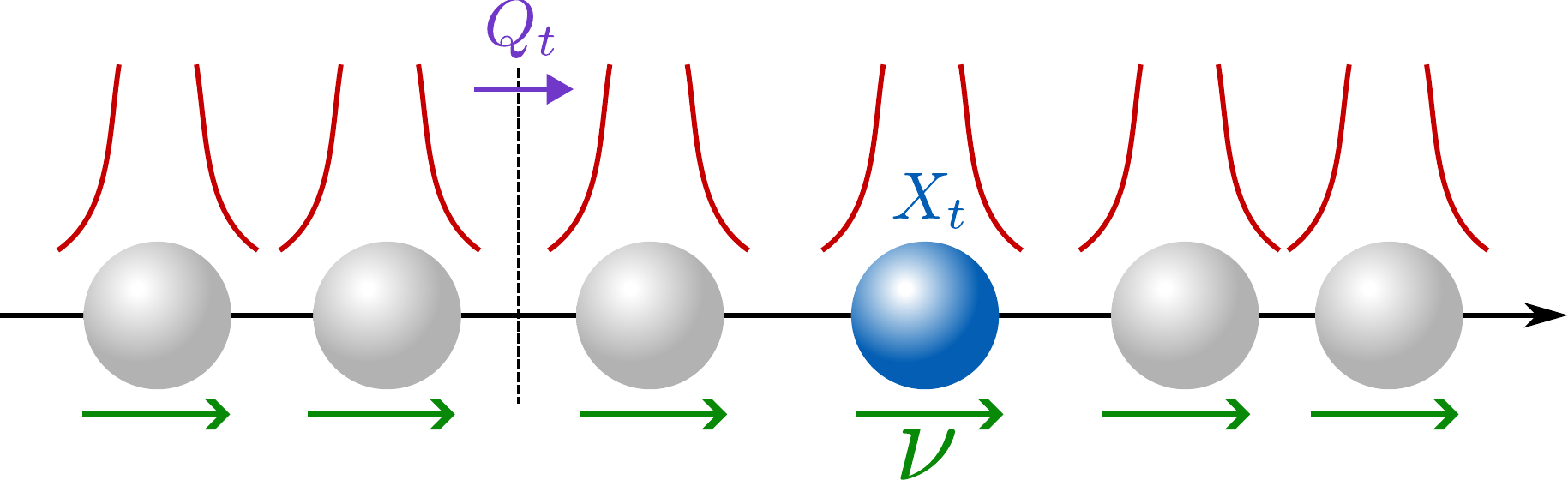}
    \caption{A system of diffusive particles in one dimension, which interact with a given potential (represented in red) and driven by a field $\nu$ (green arrows). At large scales this system can be described by two transport coefficients: the collective diffusion coefficient $D(\rho)$ and the mobility $\sigma(\rho)$~\cite{Spohn:1983,Derrida:2007,Mallick:2015,Bertini:2015}, which both depend on the local density of particles $\rho(x,t)$. We consider here two standard observables: the position $X_t$ of a tagged particle (the tracer, in blue) at time $t$, and the integrated current $Q_t$ through the origin up to time $t$.}
    \label{fig:System}
\end{figure}

A key question is how $X_T$ or $\QT$ behave when the full system is subjected to an external driving field. This is a ubiquitous situation, for instance encountered in traffic flow~\cite{Chowdhury:2000} or molecular motors~\cite{Lipowsky:2001}.  Unlike the case of non-driven single-file diffusion described above, the presence of a driving field keeps the system out of equilibrium, leading to non-trivial steady states and complex dynamics.
Despite their importance, the determination of the fluctuations (and higher order cumulants) of $X_T$ or $\QT$ for \textit{any} driven diffusive systems, i.e. the extension of~\eqref{eq:Kappa2Kollmann} and~\eqref{eq:Kappa2Qt} to driven systems, is a long standing challenge.
We stress that even in specific cases, such as the central model of the weakly asymmetric simple exclusion process (WASEP, see below for definition), the fluctuations have not been determined.
A notable exception concerns the case of Brownian particles in an external potential~\cite{Barkai:2009}.

Here, we fill this gap and characterize the fluctuations in \textit{any} driven diffusive system. 
Using the macroscopic fluctuation theory (MFT) formalism, we obtain both the second and third (which characterizes the skewness of the distribution) cumulants of the integrated current $\QT$, and the associated correlation profiles, which describe the correlations of $Q_T$ with the medium. Using a duality mapping between general 1D driven diffusive systems (which extends the one found for non-driven systems in~\cite{Rizkallah:2022}), we deduce the first three cumulants of $X_T$, as well as the bath-mediated correlations between two tracers in these systems.
We stress that our results for \textit{any} $D(\rho)$ and $\sigma(\rho)$ allow one to characterize driven systems of diffusive particles interacting with a realistic potential (such as Lennard-Jones or Weeks-Chandler-Andersen involved in single-file experiments on colloids~\cite{Wei:2000})~\footnote{If the interaction potential decays slower than $1/r$ for large distances $r$, the dynamics of the system will no longer be diffusive, but will instead depend on the precise exponent of the decay of the potential~\cite{Dandekar:2023}.}, as shown in Fig.~\ref{fig:System}.

\emph{Driven diffusive systems.---} Driven systems typically display a ballistic behaviour, meaning that the cumulants of $X_T$ or $Q_T$ grow linearly with time~\footnote{For the ASEP, under specific initial conditions one can observe a non-ballistic behaviour, such as the Kardar-Parisi-Zhang scaling $t^{1/3}$~\cite{Imamura:2012}.}. This is for instance well-known for the asymmetric simple exclusion process (ASEP), in which particles hop (with exclusion) on a 1D lattice with different rates $p_\pm$ towards the left and right~\cite{DeMasi1985,Ferrari1996,Chou:2011,Mallick:2015}.
It is a minimal model of interacting particles driven by an external field $F = (p_+ - p_-)/2$. Such systems can be studied using ballistic macroscopic fluctuation theory~\cite{Doyon:2020,Myers:2020,Perfetto:2021,Doyon:2023,Doyon:2023a}. However, for a \textit{weak drive}, the system remains diffusive, i.e. the cumulants of $Q_T$ or $X_T$ scale as $\sqrt{T}$. A typical example is the weakly asymmetric simple exclusion process (WASEP), which has attracted a lot of attention~\cite{Enaud:2004,Bodineau:2006,Bodineau:2005,prolhac2009,Gorissen:2012}, in particular due to its close relation with the Kardar-Parisi-Zhang equation~\cite{Imamura:2012}. 
In the WASEP, which is a specific case of the ASEP, the hopping rates are chosen as $p_\pm = 1 \pm \frac{\nu}{\sqrt{T}}$ when the observable ($Q_T$ or $X_T$) is measured at a large time $T$. Note that this scaling of the drive with $T$ is required to have a diffusive dynamics. 
Similarly, a general system of driven diffusive particles displays a diffusive behaviour if the drive $F$ scales with the observation time $T$ as
\begin{equation}
    \label{eq:ScalingDrive}
    F = \nu/\sqrt{T}
    \:,
    \quad
    \text{with } \nu \text{ fixed.}
\end{equation}
Importantly, this scaling of the drive with the observation time, which we take here as the definition of driven diffusive systems, allows one to obtain general results which interpolate between non-driven (equilibrium) and strongly driven (non-equilibrium) situations. Technically, we will rely on the MFT formalism.

\emph{Macroscopic fluctuation theory.} The MFT is a general framework developed by Bertini et al.~\cite{Bertini:2015} to describe diffusive systems far from equilibrium at large scales (long times and large distances). Rescaling space by $\sqrt{T}$ and time by $T$ allows to introduce a density $\rho(x,t)$ and a current $j(x,t)$ which characterise the system at the macroscopic scale. The main idea is that these fields are stochastic, and verify the constitutive equations
\begin{equation}
    \label{eq:StochDiffEq}
    \partial_t \rho +
    \partial_x j = 0
    \:,
    \quad
    j = - D(\rho) \partial_x \rho
    + \nu \sigma(\rho)
    + \frac{\sqrt{\sigma(\rho)}}{T^{1/4}} \zeta
    \:,
\end{equation}
together with the boundary conditions $\rho(\pm \infty, t) = \bar\rho$ and $j(\pm \infty, t) = \nu \sigma(\bar\rho)$, and where $\zeta$ is a Gaussian white noise, $\moy{\zeta(x,t) \zeta(x',t')} = \delta(x-x') \delta(t-t')$.
Note that the noise that arises from the stochastic dynamics scales with $T^{-1/4}$ due to the rescaling of space and time.
The transport coefficients $D(\rho)$ and $\sigma(\rho)$ embed all the microscopic details (dynamics, interaction, etc) of the system. For a given system, these transport coefficients have to be calculated from the microscopic scale. Exact formulas are available only for a few specific models, such as the WASEP for which $D(\rho) = 1$ and $\sigma(\rho) = 2 \rho(1-\rho)$.
Nevertheless, in the general case where $D(\rho)$ and $\sigma(\rho)$ cannot be computed exactly, there exists a variational formula~\cite{Spohn:1991} that yields approximate expressions~\cite{Arita:2017,Arita:2018}.

The microscopic integrated current through the origin can be expressed in terms of the macroscopic current as
\begin{equation}
\label{def:current_origin}
     \QT  = \sqrt{T} \int_{0}^{1}j(0,t) \dd t
     \:,
\end{equation}
where the factor $\sqrt{T}$ emerges from the macroscopic description (see Supplemental Material (SM)~\cite{SM} for details).
The analysis of~\eqref{def:current_origin} in the MFT framework relies on a path integral formulation of the stochastic equations~\eqref{eq:StochDiffEq}, which provides the probability of observing a trajectory $\{\rho(x,t), j(x,t)\}$. In the long times limit, only the optimal trajectory is relevant, and is determined by the Euler-Lagrange equations~\cite{Derrida:2009a,Bertini:2015},
\begin{subequations}
  \label{eq:MFTbulk}
  \begin{align}
    \partial_t q
    &= \partial_x \left[
      D(q) \partial_x q - \sigma(q) \partial_x p - \nu \sigma(q)
      \right]
      \:,
    \\
    \partial_t p
    &= - D(q)\partial_x^2 p - \frac{1}{2} \sigma'(q) (\partial_x p)^2
      - \nu \sigma'(q) \partial_x p
      \:.
  \end{align}
\end{subequations}
Here $q(x,t)$ is the optimal realisation of $\rho(x,t)$, while $p(x,t)$ is a conjugate field that imposes the conservation relation~\eqref{eq:StochDiffEq}. Roughly speaking, this field can be seen as an additional chemical potential that favors a dynamics which gives rise to an atypical fluctuation of $Q_T$.
These equations are completed by boundary conditions, which depend on the observable ($Q_T$ or $X_T$) under consideration. For $\QT$~\eqref{def:current_origin} they take the form~\cite{Derrida:2009a}
\begin{equation}
\label{eq:InitTime}
    p(x,1) = \lambda \Theta(x)
    \:, \quad
    p(x,0) =  \lambda \Theta(x) 
    + \int_{\bar\rho}^{q(x,0)} \frac{2 D(r)}{\sigma(r)} \dd r
    \:.
\end{equation}
The statistical properties of $\QT$, encoded in the cumulants $\hat{\kappa}_n$ (scaled by $\sqrt{T}$) defined by
\begin{equation}
\label{eq:psihat}
    \hat\psi(\lambda)
    \equiv \lim_{T \to \infty} \frac{1}{\sqrt{T}} \ln \big\langle \e^{\lambda Q_T} \big\rangle
    \equiv
    \sum_{n=1}^\infty \hat\kappa_n \frac{\lambda^n}{n!}
    \:,
\end{equation}
can be directly determined from the solution of the MFT equations (see SM~\cite{SM}) as
\begin{equation}
    \label{eq:CumulFromq}
    \frac{\dd \hat\psi}{\dd \lambda}
    = \nu \sigma(\rb) + \int_0^\infty [q(x,1) - q(x,0)] \dd x
    \:.
\end{equation}

Note that, even in the non-driven case $\nu = 0$, no general solution of the MFT equations~\eqref{eq:MFTbulk} is available. Exact solutions have recently been obtained~\cite{Grabsch:2022,Mallick:2022,Bettelheim:2022,Bettelheim:2022a,Krajenbrink:2022,Grabsch:2024}, but only for specific models. Here, in the case of arbitrary $D(\rho)$, $\sigma(\rho)$ and drive $\nu \neq 0$, we determine the first cumulants $\hat\kappa_n$ exactly by solving the MFT equations~(\ref{eq:MFTbulk},\ref{eq:InitTime}) order by order in $\lambda$~\cite{Krapivsky:2012,Krapivsky:2014,Krapivsky:2015a,Grabsch:2024b}.

\emph{Cumulants of the integrated current.---}  Calculations, provided in SM~\cite{SM}, finally yield explicit formulas for the first three cumulants of  $\QT$ for general  1D driven diffusive systems,
\begin{equation}
\label{kappa2}
    \hat\kappa_1 = \nu \sigma(\rb)
    \:,
    \quad
    \hat{\kappa}_2 = \frac{\sigma(\bar\rho)}{\sqrt{D(\bar\rho)}}
    \: \Fk \left( y \equiv \frac{\nu \sigma'(\rb)}{2 \sqrt{D(\rb)}}\right)
    \:,
\end{equation}
\begin{multline}
    \label{eq:Kappa3}
    \hat{\kappa}_3 = 
    \nu \frac{\sigma(\rb) \sigma'(\rb)^2}{4 D(\rb)^2}
    + \nu \frac{\sigma(\rb)^2 \sigma''(\rb) }{4 D(\rb)^2}
    \left[
    1 + 3  (y^2 - \Fk(y)^2)
    \right]
    \\
    + \frac{D'(\rb) \sigma(\rb)^2}{4 D(\rb)^{5/2}}
    \left[ 
    y (1-3 \erf(y)^2)
    - 3 \erf(y) \frac{\e^{-y^2}}{\sqrt{\pi}}
    \right]
    \:,
\end{multline}
where we denoted $\Fk(y) = \frac{\e^{-y^2}}{\sqrt{\pi}} + y \erf(y)$. For any driven diffusive system, the drive $\nu$ enters the variance $\hat{\kappa}_2$ only through the argument of the universal function $\Fk$. Note that, in general, the skewness $\hat\kappa_3 \neq 0$, even for reflecting Brownian particles ($D(\rho)=1$ and $\sigma(\rho) = 2\rho$), for which $\hat\kappa_3 = \hat\kappa_1$.
These results constitute the first description of \textit{any} driven diffusive system.
In addition, as shown below, they allow one to address important situations and go beyond the study of $Q_T$ by considering: (i) other observables such as $X_T$, (ii) the ballistic limit of driven diffusive systems, (iii) the response of the system to the perturbation induced by the displacement of the tracer, (iv) the extension to the case of several tracers. Furthermore, they also enable to reveal fundamental relation underlying the out-of-equilibrium dynamics of driven diffusive systems.

\emph{Cumulants of the tracer's position.---}  
The position $X_T$ of a tracer in a given driven diffusive system can be mapped onto the opposite of the integrated current $Q_T$ through the origin in a dual system. The main idea is that any 1D driven diffusive system can be equivalently described by a spatial density of particles, or by the distances between the particles. In this dual description, the labels of the particles play the role of their positions in the original model.
Explicitly, this can be shown (see SM~\cite{SM}) to correspond to the transformations
\begin{equation}
    \label{eq:MappingTrCoefs}
    D(\rho) \rightarrow \frac{1}{\rho^2} D \left( \frac{1}{\rho} \right)
    \:,
    \quad
    \sigma(\rho) \rightarrow \rho \: \sigma \left( \frac{1}{\rho} \right)
    \:,
\end{equation}
\begin{equation}
    \label{eq:Mappingnu}
    \nu \rightarrow-\nu
    \:,
    \quad \lambda \to - \lambda
    \:,
    \quad
    \rb \to \frac{1}{\rb}
    \:,
\end{equation}
generalising the duality relation derived in~\cite{Rizkallah:2022} for symmetric ($\nu=0$) systems. In turn, the cumulants of $X_T$ can be deduced from those of $Q_T$.
For instance the variance takes the compact form, 
\begin{equation}
    \label{eq:Kappa2tracer}
    \frac{\moy{X_T^2}_c}{\sqrt{T}}
    \underset{T \to \infty}{\simeq} 
    \frac{\sigma (\rb )}{\rb ^2 \sqrt{D(\rb )}}
    \: \Fk \left( z \equiv \nu  \frac{\sigma (\rb )-\rb  \sigma '(\rb )}{2 \rb  \sqrt{D(\rb )}} \right)
    \:,
\end{equation}
where the index $c$ refers to the cumulants. This result constitutes the extension of the celebrated result~\eqref{eq:Kappa2Kollmann} of Kollmann~\cite{Kollmann:2003} to \textit{any} driven diffusive system, in the sense that it has the same level of generality. Note that, here also, the drive $\nu$ is involved only through the same universal function $\Fk$ as for $Q_T$.
Remarkably, the effect of the drive $\nu$ on the fluctuations of $X_T$ vanishes if $\sigma(\rho) \propto \rho$. This is for instance the case in models of reflecting Brownian particles~\cite{Barkai:2009,Ryabov:2011}, for which the drive can be absorbed by a Galilean transform. In all other cases, the presence of a drive strictly increases the fluctuations of $X_T$.

\emph{Large drive limit: ballistic regime.---} Next, we show that our results, obtained under the scaling~\eqref{eq:ScalingDrive}, contain the ballistic regime as a limit. This is a priori not guaranteed, since the limits $\nu \to \infty$ and $T \to \infty$ do not necessarily commute. In fact, inserting $\nu = F \sqrt{T}$ into the cumulants~(\ref{kappa2},\ref{eq:Kappa3}) and performing the duality mapping~(\ref{eq:MappingTrCoefs},\ref{eq:Mappingnu}), we obtain
\begin{align}
    \label{eq:MeanXtBallistic}
  \moy{ X_{T}}
    &\underset{T \to \infty}{\simeq} T \: F \: \frac{\sigma(\rb)}{\rb} \:, \\
    \label{eq:VarXtBallistic}
    \moy{ X^2_{T}}_c
    &\underset{T \to \infty}{\simeq} T \: \abs{F} \: \frac{\sigma(\rb)}{2 \rb^3 D(\rb)} 
    \abs{\sigma(\rb) - \rb \sigma'(\rb)} \:,
\end{align}
\begin{multline}
    \label{eq:Kappa3XtBallistic}
    \moy{ X^3_{T}}_c
    \underset{T \to \infty}{\simeq} T \: F \: \frac{\sigma(\rb)}{4 \rb^3 D(\rb)^3} \bigg[
    \sigma''(\rb) \sigma(\rb)\\
    + \left( \sigma(\rb) - \rb \sigma'(\rb) \right) 
    \left( \frac{3 \sigma(\rb) - \rb \sigma'(\rb)}{\rb^2} + \frac{\sigma(\rb) D'(\rb)}{\rb D(\rb)} \right) \bigg].
\end{multline}
We argue that these conjectural results are the exact first cumulants of $X_T$ in the ballistic regime, for any $D(\rho)$ and $\sigma(\rho)$.
A strong indication supporting this claim is that, by setting $D(\rho) = 1$ and $\sigma(\rho)= 2 \rho(1-\rho)$, we recover from these formulas the first three cumulants of the ASEP~\cite{DeMasi1985,Ferrari1996}.
In fact, in this  case, all the cumulants are equal, $\moy{X_T^n}_c \simeq F (1-\rb)T$, indicating that the position of a tracer at long times in the ASEP is Poissonian~\cite{Ferrari1996}.
This is expected for a larger class of systems, that can be determined by imposing that the cumulants~(\ref{eq:MeanXtBallistic}-\ref{eq:Kappa3XtBallistic}) are all equal, leading to the necessary condition $D(\rho) = \frac{\abs{\sigma(\rho) - \rho \sigma'(\rho)}}{2 \rho^2}$.
Note that, considering the dual system defined by~\eqref{eq:MappingTrCoefs}, this condition simplifies into $2 D(\rho) = \abs{\sigma'(\rho)}$. In turn, this condition characterizes the celebrated zero-range processes (ZRP)~\cite{Bodineau:2004} defined as lattice models in which the hopping rates of a particle depend only on the site it currently occupies (hence ``zero range"). Finally, the statistics of $X_T$ is Poissonian only if the dual model is a ZRP.
 
\begin{figure}
    \centering
    \includegraphics[width=0.75\columnwidth]{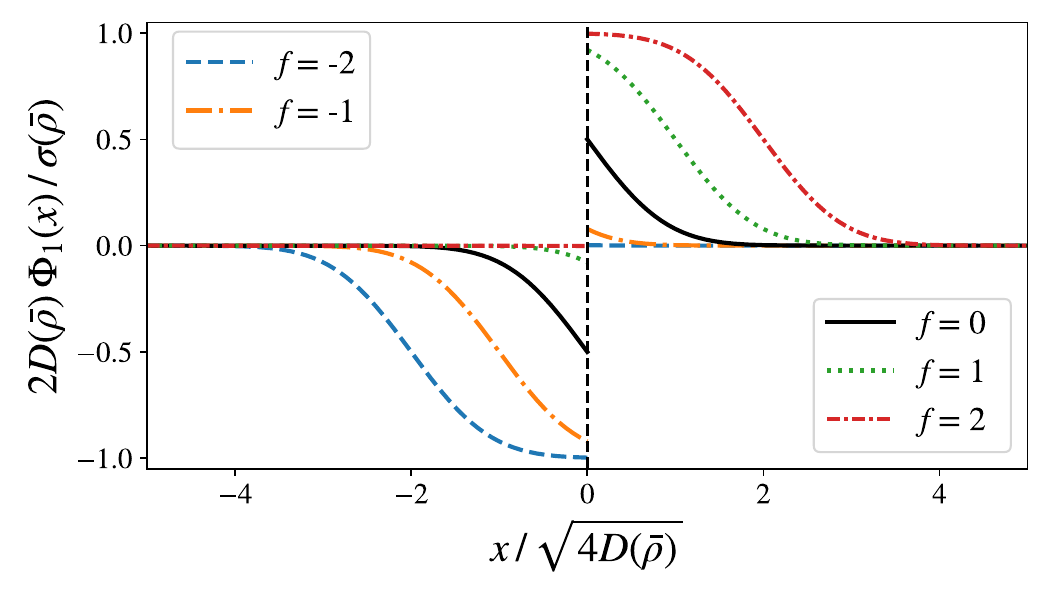}
    \caption{Correlation profile $\Phi_1(x)$~\eqref{eq:phi1} between the macroscopic density $\rho(x,1)$ (describing the microscopic system at $t=T$) and the integrated current $\QT$, as a function of $x$ for different values of the rescaled drive $f = \frac{\nu \sigma'(\rb)}{\sqrt{4 D(\rb)}}$.}
    \label{fig:CorrelProfQt}
\end{figure}

\emph{Beyond the cumulants: correlation profiles.---} Beyond the cumulants of $Q_T$ and $X_T$, our approach also gives access to the correlations between these observables and the density of surrounding particles. These correlation profiles are both (i) physically important since they quantify the response of the bath to the displacement of a tracer, and (ii) technically relevant since they were shown to satisfy a simple closed equation in the case of the SEP~\cite{Grabsch:2022}.
These correlations between the macroscopic density $\rho(x,1)$ describing the microscopic system at time $T$ and the integrated current $Q_T$ (or $X_T$) at the same time, are encoded in the generalised correlation profiles, which are given by~\cite{Poncet:2021}
\begin{equation}
    \label{def:phi}
   \frac{\moy{\rho(x,1)\e^{\lambda Q_T}}}{\moy{\e^{\lambda Q_T}}}\underset{T\to\infty}{\simeq}
   q(x,1) \equiv
   \Phi(x) = \sum_{n=0}^\infty \frac{\lambda^n}{n!} \Phi_n(x)
   \:.
\end{equation}
The correlation profiles $\Phi_n$ up to $n=2$ are determined in SM~\cite{SM}. For instance,
\begin{equation}
  \label{eq:phi1}
  \Phi_1(x)=
    \sg{x}
      \frac{\sigma(\bar\rho)}{4D(\bar\rho)} \erfc
      \left(
      \sg{x}\frac{x - \nu \sigma'(\bar{\rho})}{2\sqrt{D(\bar\rho)}}
      \right)
    \:,
\end{equation}
quantifies the correlations $\moy{\rho(x,1) Q_T}_c$ between the density and $Q_T$, and is represented in Fig.~\ref{fig:CorrelProfQt}. It is positive for $x>0$ and negative for $x<0$, indicating that an increase of $\QT$ is correlated with an accumulation of particles on $x>0$, and a depletion on $x<0$. Remarkably, when the rescaled drive $f = \frac{\nu \sigma'(\rb)}{\sqrt{4 D(\rb)}}$ is increased to a value of the order of $2$, $\QT$ completely decorrelates with the density on $x<0$ (see Fig.~\ref{fig:CorrelProfQt}). By symmetry, the same holds on $x>0$ when $f$ is decreased to the order of $-2$. Importantly, the region of space which decorrelates is not necessarily related to the direction in which the particles are driven by the field $\nu$, since $\sigma'(\rb)$ can be negative. This is for instance the case in the WASEP, for which $\sigma'(\rb) < 0$ if $\rb > \frac{1}{2}$.
We finally note that correlations between the density and the tracer, are similarly obtained by the duality mapping, and present the same qualitative behavior with $f$ replaced by $\nu\frac{\sigma(\rb)-\rb\sigma'(\rb)}{\sqrt{4 \rb^2 D(\rb)}}$.

\begin{figure}
    \centering
    \includegraphics[width=0.75\columnwidth]{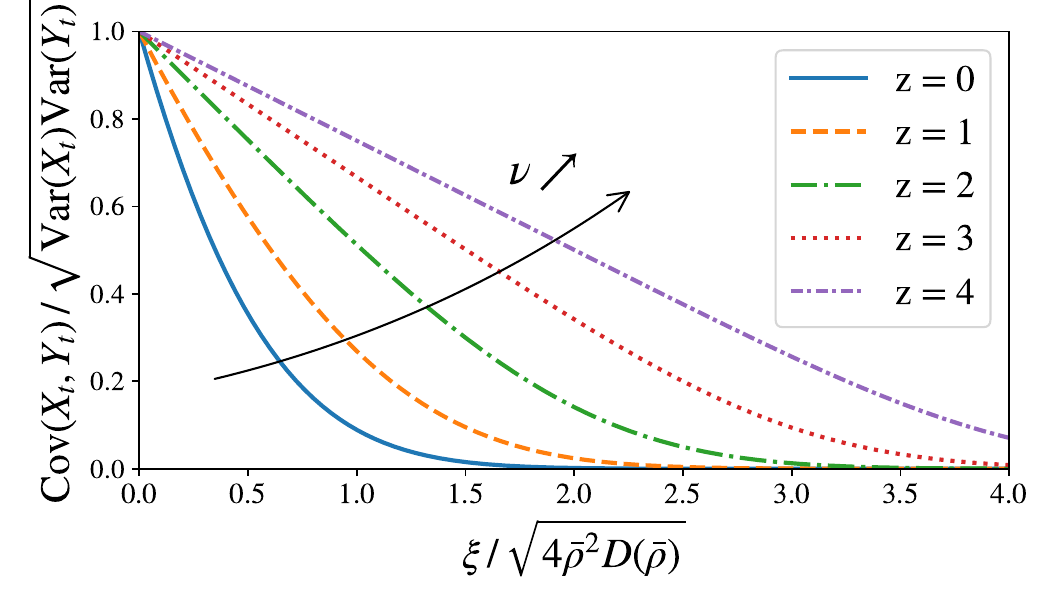}
    \caption{Normalised covariance between the positions $X_T$ and $Y_T$ of two tracers, initially separated by a distance $\xi = (Y_0 - X_0) / \sqrt{T}$. The correlations are represented for different values of $z = \nu \frac{\sigma(\bar\rho) - \bar\rho \: \sigma'(\bar\rho)}{\sqrt{4 \bar\rho^2 D(\bar\rho)}}$.}
    \label{fig:CorrelTwoTracers}
\end{figure}

\emph{Correlations between two tracers.---} Importantly, our results also allow us to address the question of the correlation between the positions of two tracers, which are expected to be strong in 1D due to bath-mediated interactions~\cite{MAJUMDAR:1991,Poncet:2018,Poncet:2019,Grabsch:2023a}.
The correlations between the positions of two tracers $X_T$ and $Y_T$, which initially start at $X_0 = 0$ and $Y_0 = \xi \sqrt{T}$~\footnote{If the two tracers are initially separated by a fixed (independent of $T$) distance, they will collapse at the macroscopic scale into a single tracer, leading to trivial correlations.} are given by (see SM~\cite{SM}),
\begin{multline}
\label{eq:variance_tracers}
   \frac{\moy{X_TY_T}_c}{\sqrt{T}}\underset{T \to +
   \infty}{\simeq} \frac{\sigma (\rb )}{\rb ^2 \sqrt{D(\rb )}}
   \Big(
    \Fk \left( z + \xi \right)
    \\
    + \Fk \left( z- \xi\right)
    -\frac{\xi }{\rb  \sqrt{D(\rb)}}
    \Big)
    \:,
\end{multline}
with $z$ defined in Eq.~\eqref{eq:Kappa2tracer}. Note that this covariance involves again the same universal function $\Fk$ as for the variance of $Q_T$ and $X_T$.
As expected, these correlations decrease when the initial distance between the tracer increases, and increase with the drive (see Fig.~\ref{fig:CorrelTwoTracers}). Our Eq.~\eqref{eq:variance_tracers} quantifies these trends for \textit{any} driven diffusive system.

\emph{Physical boundary conditions for driven diffusive systems.---} In addition to providing answers to the variety of questions considered above, our approach also enables us to reveal fundamental relations underlying the out-of-equilibrium dynamics of driven diffusive systems.
Indeed, we claim that the profile $\Phi$~\eqref{def:phi} satisfies the set of simple physical boundary conditions,
\begin{equation}
    \label{eq:BoundCondPhi}
   \mu(\Phi(0^+)) - \mu(\Phi(0^-)) = \lambda
    \:,
    \quad
    \left[ \partial_z \mu(\Phi) \right]_{0^-}^{0^+}
    = 0
    \:,
\end{equation}
\begin{equation}
    \label{eq:short_cut}
    \hat{\psi}+\nu \partial_\nu\hat{\psi} = 
    -2\left(\partial_x \mu(\Phi)|_{0^+}-2\nu\right)
    \int^{\Phi(0^+)}_{\Phi(0^-)}D(r)\dd r 
    \:,
\end{equation}
where $\mu(\rho) = \int^{\rho} \frac{2 D(r)}{\sigma(r)} \dd r$ is the chemical potential and $[f]_{a}^{b} = f(b) - f(a)$. These relations also hold for the profiles associated to $X_T$, but with $\mu$ replaced by the pressure ${P}({\rho}) = \int^{\rho} \frac{2 r {D(r)}}{{\sigma}(r)} \dd r$ in the first relation of~\eqref{eq:BoundCondPhi}.

The conditions~\eqref{eq:BoundCondPhi} are obtained from the MFT equation~(\ref{eq:MFTbulk},\ref{eq:InitTime}). Remarkably, these relations have the same form as in the non-driven case~\cite{Grabsch:2024}, which in turn demonstrates their generality.  Equation~\eqref{eq:short_cut} generalizes the boundary condition derived for symmetric  ($\nu=0$) systems~\cite{Grabsch:2024b} to general driven diffusive systems and is a conjecture, supported by the following points: (i) Based on our determination of the first three cumulants~(\ref{kappa2},\ref{eq:Kappa3}), we verify that it holds for arbitrary $D(\rho), \sigma(\rho)$ and $\nu$ up to order 3 in $\lambda$. (ii) For arbitrary $\lambda$, this relation is verified in the case of the WASEP (see SM~\cite{SM}). (iii) Relation~\eqref{eq:short_cut} is invariant under the duality mapping~(\ref{eq:MappingTrCoefs},\ref{eq:Mappingnu}).
Technically, this last relation provides a shortcut to the determination of the cumulants from the correlation profile $\Phi$ as it bypasses the computation of the integral~\eqref{eq:CumulFromq}.

We stress that the boundary relations~(\ref{eq:BoundCondPhi},\ref{eq:short_cut}) hold in out-of-equilibrium situations even if they involve thermodynamic functions, defined for equilibrium systems. The key point is that these functions are evaluated for the out-of-equilibrium quantities $\Phi(0^\pm)$ and $\Phi'(0^\pm)$.

\emph{Conclusion.---}
We have considered both the integrated current of particles $Q_T$ and the displacement $X_T$ of a tracer in \textit{any} driven diffusive system. We have determined explicit expressions for the first three cumulants of both quantities, and in particular provided the extension of the celebrated result of Kollmann~\cite{Kollmann:2003} to any driven diffusive system. In addition to characterizing the fluctuations of these observables and the skewness of their distributions, we have characterized the response of the bath by determining (i) the full spatial structure of the bath-observable correlations (up to second order) and (ii) the correlations between the positions of two tracers, mediated by the bath.
Importantly, we have also obtained general physical boundary conditions satisfied by the correlation profiles in these out-of-equilibrium systems.
Altogether, this work constitutes a step toward the determination of the full distribution of all these observables in driven one-dimensional systems of interacting particles.

\emph{Acknowledgements.---} The authors thank Kirone Mallick, Alexandre Krajenbrink and Pierre Le Doussal for inspiring discussions.\\

\bibliographystyle{apsrev4-2}

\clearpage
\widetext

\let\addcontentsline\oldaddcontentsline

\begin{center}
  \begin{large}

    \textbf{
     Supplemental Material for\texorpdfstring{\\}{} Tracer and current fluctuations in driven diffusive systems
   }
  \end{large}
   \bigskip

    Théotim Berlioz, Olivier Bénichou and  Aurélien Grabsch 
\end{center}

\setcounter{equation}{0}
\makeatletter

\renewcommand{\theequation}{S\arabic{equation}}
\renewcommand{\thefigure}{S\arabic{figure}}

\renewcommand{\bibnumfmt}[1]{[S#1]}
\renewcommand{\citenumfont}[1]{S#1}

\setcounter{secnumdepth}{3}

\tableofcontents

\section{Microscopic to macroscopic description}
\label{sec:MicroToMacro}

We first show how we can build a macroscopic description at a long time $T$ from a microscopic model of interacting particles, and in particular how the integrated current at the microscopic scale is related to the macroscopic fields. We discuss the two cases of lattice gases (particles hop on discrete sites) and of particles on the real line.

\subsection{Lattice gases}

We first consider particles on a 1D infinite lattice, and denote $n_i(t)$ the number of particles at site $i$ at time $t$. In exclusion models such as the SEP or the ASEP, $n_i$ can only take values $0$ or $1$, but other models such as zero range processes can host an arbitrary number of particles per site~\cite{Evans:2005SM}. We also introduce $Q_t^{(i)}$ which counts the number of particles that have crossed the bond from $i$ to $i+1$, minus those that have crossed from $i+1$ to $i$, up to time $t$. This is the integrated current up to $t$ on each site $i$. These quantities satisfy a microscopic conservation relation,
\begin{equation}
    \label{eq:MicroCons}
    n_i(t) - n_i(0) = Q_t^{(i-1)} - Q_t^{(i)}
    \:,
\end{equation}
which states that the variation of the number of particle on site $i$ comes from the particles that have arrived from site $i-1$ minus those which have left to site $i+1$.

We construct a macroscopic description of this problem by rescaling time by the large observation time $T$ and space by $\sqrt{T}$ (since we consider systems with diffusive dynamics). We introduce macroscopic density and current fields, defined as~\cite{Derrida:2007SM,Bertini:2015SM},
\begin{equation}
    \label{eq:DefMacroFieldsFromLattice}
    \rho(x,t) = \frac{1}{\sqrt{T}}
    \sum_{i} n_i(t T) \delta \left( x - \frac{i}{\sqrt{T}} \right)
    \:,
    \quad
    j(x,t) =
    \sum_{i} \delta \left( x - \frac{i}{\sqrt{T}} \right)
    \left. \dt{}{t} Q_t^{(i)} \right|_{t \to t T}
    \:,
\end{equation}
where we have denoted $\delta(x)$, the Dirac delta function. From the microscopic conservation equation~\eqref{eq:MicroCons}, we obtain that $\rho$ and $j$ must satisfy
\begin{equation}
    \label{eq:ConsMacroV1}
    \partial_t \rho =
    \sqrt{T} \left[ 
    j \left( x - \frac{1}{\sqrt{T}}, t \right)
    - j(x,t)
    \right]
    \:.
\end{equation}
If instead of $\delta$-functions we use smooth smearing functions (normalised to $1$), $j$ is continuous so that~\eqref{eq:ConsMacroV1} becomes the usual continuity equation,
\begin{equation}
    \partial_t \rho + \partial_x j = 0
    \:,
\end{equation}
for large $T$. From the definition of $j$~\eqref{eq:DefMacroFieldsFromLattice}, we can express the integrated current through the origin as
\begin{equation}
    Q_t \equiv
    Q_t^{(0)}
    = T \int_{-\frac{1}{\sqrt{2T}}}^{\frac{1}{2\sqrt{T}}} \dd x'
    \int_0^{t/T} \dd t' \: j \left( x', t' \right)
    \:.
\end{equation}
Replacing again $j$ by a smooth function, this becomes for large $T$,
\begin{equation}
    Q_{T} = \sqrt{T} \int_0^1 j(0,t') \dd t'
    \:,
\end{equation}
which corresponds to Eq.~(5) in the main text.

Similarly, we define the generating function of the correlations between the occupation numbers and the current $Q_T$ as
\begin{equation}
    \label{eq:DefCorrelMicro}
    \sum_{n = 0}^\infty \frac{\lambda^n}{n!} \moy{n_i(T) Q_T^n}_c
    =
    \frac{\moy{n_i(T) \: \e^{\lambda Q_T}}}{\moy{\e^{\lambda Q_T}}}
    = \frac{\moy{ \rho(x,1) \e^{\lambda Q_T} }}{\moy{\e^{\lambda Q_T}}}
    \:,
    \quad
    x = \frac{i}{\sqrt{T}}
    \:,
\end{equation}
which is Eq.~(18) in the main text.

\subsection{Continuous models}

Let us consider now a model of particles on the real line, at positions $x_n(t)$ at time $t$, with $n \in \mathbb{Z}$ the label of the particle. We can construct density and current fields at the microscopic level,
\begin{equation}
    \rhom(x,t) =
    \sum_n \delta(x - x_n(t))
    \:,
    \quad
    \jm(x,t)
    = \sum_n  \delta \big (x - x_n(t) \big) \dt{}{t} x_n(t)
    \:,
\end{equation}
which already verify the continuity equation at the microscopic scale,
\begin{equation}
    \partial_t \rhom + \partial_x \jm = 0
    \:.
\end{equation}
The macroscopic fields are defined by rescaling space by $\sqrt{T}$ and time by $T$,
\begin{equation}
    \rho(x,t) \equiv \rhom (  x \sqrt{T}, tT )
    \:,
    \quad
    j(x,t) \equiv \sqrt{T} \: \jm \left( x \sqrt{T}, t T \right)
    \:,
\end{equation}
where the factor $\sqrt{T}$ in the current is placed to ensure the continuity relation
\begin{equation}
    \partial_t \rho + \partial_x j = 0
    \:.
\end{equation}
In particular, it indicates implies that the microscopic integrated current though the origin up to $T$ takes the form
\begin{equation}
    Q_T \equiv \int_0^T \jm(0,t') \dd t'
    = \frac{1}{\sqrt{T}} \int_0^T j \left( 0, \frac{t'}{T} \right) \dd t'
    = \sqrt{T} \int_0^1 j(0,t') \dd t'
    \:,
\end{equation}
which is again Eq.~(5) in the main text.

We also define the generating function of the correlations between the microscopic density and the integrated current $Q_T$ as
\begin{equation}
    \sum_{n = 0}^\infty \frac{\lambda^n}{n!} \moy{\rhom(y,T) Q_T^n}_c
    =
    \frac{\moy{\rhom(y,T) \: \e^{\lambda Q_T}}}{\moy{\e^{\lambda Q_T}}}
    = \frac{\moy{ \rho(x,1) \e^{\lambda Q_T} }}{\moy{\e^{\lambda Q_T}}}
    \:,
    \quad
    x = \frac{y}{\sqrt{T}}
    \:,
\end{equation}
which is again Eq.~(18) in the main text.

\section{Macroscopic fluctuation theory}

In this Section, we briefly summarise the main tools needed to study both the current and tracer properties of different single-file systems.

\subsection{The transport coefficients and the fluctuating hydrodynamics}
\label{sec:FlucHydro}

Bertini et al. \cite{Bertini:2015SM} introduced a hydrodynamic description for out-of-equilibrium (both driven and non-driven) microscopic diffusive systems. This approach asserts that at large times and distances, the system is fully characterized by the coarse-grained density $\rho(x,t)$ and the local current $j(x,t)$ (which can be defined for lattice gases or continuous models, see Section~\ref{sec:MicroToMacro}). Under local equilibrium, their evolution is governed by the macroscopic conservation equation and Fick's law:
\begin{align}
\label{conservation}
     &\partial_t \rho = -\partial_x j \,\\
\label{current_macro_determ}
     &j = -D(\rho) \partial_x \rho + \nu \sigma(\rho) 
\end{align}
where the transport coefficient have been noted $D(\rho)$ for the diffusivity and $\sigma(\rho)$ for the mobility (sometimes called conductivity) and $\nu$ is the driving force. The diffusivity, which depends in general of the local density, characterizes the relaxation time of the density. Similarly to the diffusivity, the mobility $\sigma(\rho)$ is a function of the local density, quantifying the response to an external field.
We stress that, at the macroscopic level, all the microscopic details are encapsulated into these two transport coefficients $D(\rho)$ and $\sigma(\rho)$. Initially defined for a lattice gas~\cite{Spohn:1983SM}, they can be more intuitively understood~\cite{Mallick:2015SM} for any system by considering a finite system of length $L$ between two reservoirs with densities $\rho_{\mathrm{L}}$ and $\rho_{\mathrm{R}}$~\cite{Derrida:2007SM}. Let $Q_t$ represent the total number of particles transferred from the left reservoir to the right one up to time $t$. The diffusion coefficient then measures the average current in the presence of a small density difference, and the mobility $\sigma(\rho)$ the response to the driving force $\nu$,
\begin{equation}
    \label{eq:DefDsigmaFromQt}
    \lim_{t \to \infty} \frac{\moy{Q_t}}{t} = 
    -\frac{D(\rho)}{L} (\rho_{\mathrm{R}} - \rho_{\mathrm{L}})+\nu \sigma(\rho)
    \:,
    \quad
    \text{for}
    \quad
    \rho_{\mathrm{R}} - \rho_{\mathrm{L}} \ll 
    \rho \equiv \frac{\rho_{\mathrm{R}} + \rho_{\mathrm{L}}}{2}
    \:.
\end{equation}
Under local equilibrium, these two quantities are linked by the Einstein relation, a result derived from the fluctuation-dissipation theorem, which is verified at equilibrium~\cite{Bertini:2015SM}
\begin{equation}
    \frac{2D(\rho)}{\sigma(\rho)} = \frac{d^2 f(\rho)}{d\rho^2},
\end{equation}
where $f(\rho) = \int^\rho \mu(r) \dd r$ is the density free-energy at equilibrium and $\mu$ the chemical potential. 

Importantly, the mobility also controls the fluctuations of the current at equilibrium,
\begin{equation}
    \label{eq:DefDsigmaFromFlucQt}
    \lim_{t \to \infty} \frac{\moy{Q_t^2}}{t} = 
    \frac{\sigma(\rho)}{L}
    \:,
    \quad
    \text{for}
    \quad
    \rho = \rho_{\mathrm{R}} = \rho_{\mathrm{L}} \ \text{and} \ \nu =0
    \:.
\end{equation}
The main idea of the macroscopic fluctuation theory is to treat $\rho$ and $j$ as stochastic fields, which still verify the continuity equation~\eqref{conservation}, but completed by a stochastic version of the Fick law~\eqref{current_macro_determ},
\begin{equation}
    \label{current stochastic_micro}
    j(x,t) =- D(\rho(x,t)) \partial_x \rho(x,t) + \nu \sigma(\rho(x,t))+\sqrt{\frac{\sigma(\rho(x,t))}{\sqrt{T}}}\zeta(x,t) \:,
\end{equation}
where $\zeta$ is a Gaussian white noise delta correlated in space and time, meaning $\moy{\zeta(x,t) \zeta(x',t')} = \delta(x-x') \delta(t-t')$.
Note that the noise~\eqref{current stochastic_micro} scales as $T^{-1/4}$ due to the rescaling of space and time performed to construct $\rho$ and $j$~\cite{Bertini:2015SM}.
Combining equations \eqref{conservation} and \eqref{current stochastic_micro}, we obtain the stochastic equation satisfied by the density $\rho$ (we recall that $\rho$ describes the system at the macroscopic scale, see Section~\ref{sec:MicroToMacro}),
\begin{equation}
    \label{eq:StochDiff}
    \partial_t \rho = \partial_x \left[
    D(\rho) \partial_x \rho
    - \nu \sigma(\rho)+\sqrt{\frac{\sigma(\rho)}{\sqrt{T}}} \zeta
    \right]
    \:.
\end{equation}

\subsection{The current through the origin}

As we have seen in Section~\ref{sec:MicroToMacro}, the integrated current through the origin \( \Qt \) is expressed in terms of the macroscopic current $j$ as
\begin{equation}
  \label{eq:RelIntegrCurr}
  \Qt  = \sqrt{T}\int_0^{t/T} \dd t' j(0,t')
  \:.
\end{equation}
As we will see below, the MFT formalism is more conveniently implemented by expressing $\Qt$ in terms of a different current
\begin{equation}
    \label{eq:DefQtilde}
    \tilde{Q}_t \equiv 
    \int_0^\infty\left(\rho(x,t)-\rho(x,0)\right) \dd x 
    \:,
\end{equation}
which has the advantage to involve only the density $\rho$ at two different times ($t=0$ and the macroscopic measurement time $t$). The two expressions~(\ref{eq:RelIntegrCurr},\ref{eq:DefQtilde}) are related by the continuity equation~\eqref{conservation}. This can be shown by taking the time derivative of~\eqref{eq:DefQtilde},
\begin{equation}
  \frac{\dd \tilde{Q}_t}{\dd t} =
  \int_{0}^\infty \partial_t \rho(x,t) \, \dd x
  = -  \int_0^\infty \partial_x j(x,t) \, \dd x
  \:.
\end{equation}
Using that \( \rho(x,t) \to \rb \) when \( x \to \infty \) and $j(x,t) \to \nu \sigma(\rb)$, we get:
\begin{equation}
  \dt{\tilde{Q}_t}{t} = j(0,t) - j(+\infty,t)  =j(0,t)- \nu \sigma(\rb)
  \:.
\end{equation}
Integrating up to time $t/T$, and combining with (\ref{eq:RelIntegrCurr}), we get:
\begin{equation}
  \label{eq:CurrentASEP}
    \Qt =  \sqrt{T} \int_0^{t/T} j(0,t')\dd t' =
    \sqrt{T} \int_0^{t/T} \nu \sigma(\rb)\dd t'
    + \sqrt{T} \: \tilde{Q}_{t/T}
    =
    \nu \sigma(\rb) \frac{t}{\sqrt{T} }
    + \sqrt{T}  \: \tilde{Q}_{t/T}
    \:.
\end{equation}
Setting $t = T$, we obtain
\begin{equation}
  \label{eq:Currentrescaledrelation}
    \QT
    = \sqrt{T} \left(\nu \sigma(\rb) 
    + \tilde{Q}_1 \right)
    = \sqrt{T} \left(\nu \sigma(\rb) 
    + \int_0^\infty\left(\rho(x,1)-\rho(x,0)\right) \dd x\right)
    \:.
\end{equation}

\subsection{MFT for the integrated current}

We sketch here the main steps of the derivation of the MFT equations for the integrated  current $Q_T$. From~\eqref{eq:Currentrescaledrelation}, the integrated current can be written as a functional of the density as
\begin{equation}
    \label{eq:DefHatQ}
    Q_T = \sqrt{T} \hat{Q}[\rho]
    \:,
    \quad
    \hat{Q}[\rho]
    \equiv
    \nu \sigma(\rb) 
    + \int_0^\infty\left(\rho(x,1)-\rho(x,0)\right) \dd x
    \:.
\end{equation}
We can thus write the moment generating function of $Q_T$~\eqref{eq:RelIntegrCurr} using a path integral formulation~\cite{Derrida:2009aSM},
\begin{equation}
    \label{eq:MomGenFctMFT}
   \moy{\e^{\lambda Q_T}}
    = \int \mathcal{D} \rho(x,t) \mathcal{D} H(x,t) 
    \int \mathcal{D} \rho(x,0) \:
    \e^{- \sqrt{T}(S[\rho,H] + F[\rho(x,0)]-\lambda \hat{Q}[\rho])}
    \:,
\end{equation}
which is a reformulation of the stochastic equation~\eqref{eq:StochDiff}. We have denoted $S$ the MFT action
\begin{equation}
    \label{eq:MFTaction}
    S[\rho,H] = \int_{-\infty}^\infty \dd x \int_0^T \dd t \left[
    H \partial_t \rho
    + D(\rho) \partial_x \rho \partial_x H
    - \frac{\sigma(\rho)}{2} (\partial_x H)^2- \nu \sigma(\rho) \partial_x H
    \right]
    \:,
\end{equation}
$F$ the free-energy, which gives the distribution of the (annealed) initial condition picked from a steady state density $\rb$,
\begin{equation}
    F[\rho(x,0)] = \int_{-\infty}^\infty \dd x
    \int_{\rb}^{\rho(x,0)} \dd r
    \left[ \rho(x,0) - r \right] \frac{2 D(r)}{\sigma(r)}
    \:,
\end{equation}
and $H$ a Lagrange multiplier that enforces the local conservation of particles~\eqref{conservation}.
Thanks to the factor $\sqrt{T}$ in the exponential in~\eqref{eq:MomGenFctMFT}, the functional integrals can be evaluated by a saddle point method in the large time limit. Let us denote $(q,p)$ the fields $(\rho,H)$ which minimize $S + F - \lambda \hat{Q}$. They satisfy the MFT equations~\cite{Derrida:2009aSM}
\begin{align}
  \label{eq:MFT_q}
  \partial_t q &= \partial_x[D(q) \partial_x q
  - \sigma(q)\partial_x p +\nu\sigma(q)]
  \:,
  \\
  \label{eq:MFT_p}
  \partial_t p &= - D(q) \partial_x^2 p 
  - \frac{1}{2}  \sigma(q) (\partial_x p)^2 -\nu\sigma'(q)\partial_x p
  \:,
\end{align}
with the final and initial conditions
\begin{equation}
\label{eq:InitTimeSM}
    p(x,1) = \lambda \Theta(x)
    \:, \quad
    p(x,0) =  \lambda \Theta(x) 
    + \int_{\bar\rho}^{q(x,0)} \frac{2 D(r)}{\sigma(r)} \dd r
    \:.
\end{equation}
In terms of the saddle point solution $(q,p)$, the cumulant generating function is directly obtained from~\eqref{eq:MomGenFctMFT} as
\begin{equation}
    \label{eq:PsiFromMFT}
    \ln \moy{\e^{\lambda Q_T}}
    \underset{T \to \infty}{\simeq}
    \sqrt{T} \hat\psi 
    \:,
    \quad 
    \hat\psi \equiv \lambda \hat{Q}[q] - F[q(x,0)] - S[q,p]
    \:.
\end{equation}
In practice, this relation is rather difficult to use since it requires to compute a space-time integral for $S$~\eqref{eq:MFTaction}, which involves the solution at all times. It is much simpler to compute the derivative of the cumulant generating function with respect to $\lambda$, which reduces to
\begin{equation}
    \label{eq:dpsidlambda}
    \dt{\hat\psi}{\lambda}
    = \hat{Q}[q]
    \:,
\end{equation}
since $(q,p)$ is the minimum of $S + F - \lambda \hat{Q}$. Since $\hat{Q}$ is given by~\eqref{eq:DefHatQ}, it only involves the optimal field $q$ at initial time $t=0$ and final time $t=1$. We will use this relation to determine the first cumulants of $Q_T$ in the following.

\subsection{Duality mapping}
\label{subsection:duality}

Recently, a general mapping between symmetric diffusive single-file systems has been discovered. It maps the position of a tracer in a single-file system onto the integrated current through the origin in a dual model~\cite{Rizkallah:2022SM}. In this section, we extend this mapping to driven diffusive systems.

The general idea of this duality mapping is that the same model can be described in two equivalent ways, by looking either at the positions of the particles, or at the distance between two consecutive particles. At the macroscopic level, this duality becomes a mapping between the stochastic equations~\eqref{eq:StochDiff}. Let us first consider a model described by the density $\rho$ and current $j$ which obey
\begin{equation}
    \label{eq:StochEqOriginal}
    \partial_t  \rho + \partial_x j = 0
    \:, 
    \quad
    j=-D(\rho) \partial_x \rho
    + \nu \sigma(\rho)+\frac{\sqrt{\sigma(\rho)}}{T^{1/4}} \zeta
    \:,
\end{equation}
with $\zeta$ a Gaussian white noise, delta correlated in space and time. We introduce dual density and current fields $\tilde{\rho}$ and $\tilde{j}$, defined from the original ones as
\begin{equation}
    \label{eq:MappingDens}
   \rho(x,t) = \frac{1}{\tilde{\rho}(k(x,t),t)}
    \:,
    \quad
    j(x,t) = -\frac{\tilde{j}(k(x,t), t)}{\tilde{\rho}(k(x,t), t)}
    \:,
    \quad
    \partial_x k(x,t) = \rho(x,t)
    \:,
    \quad
    \partial_t k(x,t) = -j(x,t)
    \:.
\end{equation}
Inserting this transformation into the stochastic equations~\eqref{eq:StochEqOriginal}, we obtain that $\tilde{\rho}$ and $\tilde{j}$ obey the same equation,
\begin{equation}
    \label{eq:StochEqDual}
    \partial_t  \tilde\rho + \partial_k \tilde{j} = 0
    \:, 
    \quad
    \tilde{j}=-\tilde{D}(\tilde\rho) \partial_k \tilde\rho
    + \tilde\nu \tilde\sigma(\tilde\rho)
    +\frac{\sqrt{\tilde\sigma(\tilde\rho)}}{T^{1/4}} \zeta
    \:,
\end{equation}
with the new dual transport coefficients,
\begin{equation}
    \label{eq:MappingTrCoefsSM}
    \tilde{D}(\rho) = \frac{1}{\rho^2} D \left( \frac{1}{\rho} \right)
    \:,
    \quad
    \tilde\sigma(\rho) = \rho \: \sigma \left( \frac{1}{\rho} \right)
    \:,
    \quad
   \tilde\nu=-\nu
    \:.
\end{equation}
Note that this relies on the fact that $\zeta$ is a Gaussian white noise, so that  \( \zeta(x,t) \eqlaw - \zeta(x,t) \), and
\begin{equation}
    \zeta(k(x,t),t) \eqlaw \sqrt{\dep{x}{k}} \: \zeta(x,t)
    = \frac{1}{\sqrt{\rho(x,t)}} \zeta(x,t)
    \:.
\end{equation}
Due to the transformation of the density~\eqref{eq:MappingDens}, the mean density $\rb$ of the original model becomes $\tilde{\bar\rho} \equiv \frac{1}{\rb}$ for the dual model.

\bigskip

Importantly, $k(x,t)$ can be interpreted as the label of the particle located at position $x$ at time $t$ (at the macroscopic scale, the labels become continuous due to the rescaling by $\sqrt{T}$). Therefore, inverting the transformation~\eqref{eq:MappingDens} gives the position of the $k^{\mathrm{th}}$ particle at time $t$ as
\begin{equation}
    \label{eq:MappingCurrTracGen}
    x(k,t) - x(k,0) = \int_0^t \tilde{j}(k,t') \dd t'
    \:.
\end{equation}
This relation indicates that the displacement of the $k^{\mathrm{th}}$ particle in the original system described by~\eqref{eq:StochEqOriginal} actually corresponds to the opposite of the integrated current through the position $k$ in the dual model described by~\eqref{eq:StochEqDual}, so that
\begin{equation}
    \ln \moy{\e^{\lambda X_k(t)}} \Big|_{D,\sigma} = 
    \ln \moy{\e^{- \lambda Q_t(k)}} \Big|_{\tilde{D},\tilde\sigma}
    \:,
\end{equation}
with $X_k$ the displacement of the $k^{\mathrm{th}}$ particle, and $Q_t(k)$ the integrated current through $k$ in the dual model. Expanding this relation in powers of $\lambda$, we can directly map the cumulants of the integrated current onto those of a tracer by performing the substitution~\eqref{eq:MappingTrCoefsSM}. This will be useful below.

\section{Solving the MFT equations for the first cumulants}
\label{sec:PertSol}

We will solve the MFT equations~(\ref{eq:MFT_q},\ref{eq:MFT_p})
in powers of $\lambda$ to obtain the first cumulants and associated correlation profiles. Because of the boundary
conditions~(\ref{eq:InitTimeSM}), $p$ is of order $\lambda$. We
define the expansions
\begin{equation}
  \label{eq:DefExppq}
  q = \bar\rho + \lambda q_1 + \lambda^2 q_2 + \mathcal{O}(\lambda^3)
  \:,
  \quad
  p = \lambda p_1 + \lambda^2 p_2 + \mathcal{O}(\lambda^3)
  \:.
\end{equation}

When using these expansions in the bulk equations~(\ref{eq:MFT_q},\ref{eq:MFT_p}), we get a drift term \( -\nu \sigma'(\bar{\rho}) \partial_x \) in both equations. We can absorb this term by defining the fields in a moving frame,
\begin{equation}
  \label{eq:ChangeFrame}
  q_n(x,t) = Q_n(x - \nu \sigma'(\bar\rho) t, t)
  \:,
  \quad
  p_n(x,t) = P_n(x - \nu \sigma'(\bar\rho) t, t)
  \:.
\end{equation}

\subsection{First Order}
At first order in $\lambda$, the MFT equations become
\begin{equation}
  \partial_t P_1 = - D(\bar\rho)\partial_x^2 P_1
  \:,
  \quad
  \partial_t Q_1 = D(\bar\rho)\partial_x^2 Q_1 - \sigma(\bar\rho) \partial_x^2 P_1
  \:,
\end{equation}
with the boundary conditions~(\ref{eq:InitTimeSM}) which yield
\begin{equation}
  P_1(x,1) = \Theta(x + \nu \sigma'(\bar\rho))
  \:,
  \quad
  Q_1(x,0) = \frac{\sigma(\bar\rho)}{2D(\bar\rho)} (P_1(x,0) - \Theta(x))
  \:.
\end{equation}
The equation on $P_1$ gives straightforwardly
\begin{equation}
  P_1(x,t) = \frac{1}{2} \erfc \left(- \frac{x + \nu \sigma'(\bar{\rho})}{2 \sqrt{D(\bar\rho)(1-t)}} \right)
  \:.
\end{equation}
For $Q_1$ it is convenient to define
\begin{equation}
  Q_1(x,t) = \frac{\sigma(\bar\rho)}{2D(\bar\rho)} (P_1(x,t) + \tilde{Q}_1(x,t))
  \:,
\end{equation}
so that
\begin{equation}
  \partial_t \tilde{Q}_1 = D(\bar\rho)\partial_x^2 \tilde{Q}_1
  \:,
  \quad
  \tilde{Q}_1(x,0) = - \Theta(x)
  \:.
\end{equation}
We thus get
\begin{equation}
  \tilde{Q}_1(x,t) = - \frac{1}{2} \erfc \left( - \frac{x}{2\sqrt{D(\bar\rho)t}} \right)
  \:.
\end{equation}
In particular, we get from these expressions
\begin{equation}
  \label{q1T1}
  q_1(x,1) =
  \left\lbrace
    \begin{array}{ll}
      \displaystyle
      \frac{\sigma(\bar\rho)}{4D(\bar\rho)} \erfc
      \left(
      \frac{x - \nu \sigma'(\bar{\rho})}{2\sqrt{D(\bar\rho)}}
      \right)
      &
        \text{for } x > 0
        \:,
      \\[0.4cm]
      \displaystyle
      -\frac{\sigma(\bar\rho)}{4D(\bar\rho)} \erfc
      \left(
      -\frac{x - \nu \sigma'(\bar{\rho})}{2\sqrt{D(\bar\rho)}}
      \right)
      & \text{for } x < 0
        \:,
    \end{array}
  \right.
\end{equation}
\begin{equation}
\label{q1T0}
  q_1(x,0) =
  \left\lbrace
    \begin{array}{ll}
      \displaystyle
      -\frac{\sigma(\bar\rho)}{4D(\bar\rho)} \erfc
      \left(
      \frac{x + \nu \sigma'(\bar{\rho})}{2\sqrt{D(\bar\rho)}}
      \right)
      
      &
        \text{for } x > 0
        \:,
      \\[0.4cm]
      \displaystyle
      \frac{\sigma(\bar\rho)}{4D(\bar\rho)} \erfc
      \left(
      -\frac{x + \nu \sigma'(\bar{\rho})}{2\sqrt{D(\bar\rho)}}
      \right)
      & \text{for } x < 0
        \:.
    \end{array}
  \right.
\end{equation}

\subsection{Second order}

At second order in $\lambda$, the MFT equations equations are:
\begin{equation}
\partial_t P_2 = - D(\bar\rho)\partial_x^2 P_2-\frac{1}{2}\sigma'(\bar\rho)(\partial_x P_1)^{2}-\left(\nu \sigma''(\bar\rho)\partial_x P_1+D'(\bar\rho)\partial_x^{2} P_1\right)Q_1,$$
  $$\partial_t Q_2 = D(\bar\rho)\partial_x^2 Q_2 -\sigma'(\bar\rho)\partial_x P_1\partial_x Q_1-\sigma(\bar\rho)\partial_x^2 P_2-\left(\nu \sigma''(\bar\rho)\partial_x Q_1+\sigma'(\bar\rho)\partial_x^{2} P_1-D'(\bar\rho)\partial_x^{2} Q_1\right)Q_1+D'(\bar\rho)(\partial_x Q_1)^{2},
\end{equation}
with the boundary conditions~(\ref{eq:InitTimeSM}) which yield
\begin{equation}
\label{boundary_order2_Arb}
  P_2(x,1) = 0
  \:,
  \quad
  Q_2(x,0) = \frac{\sigma(\bar\rho)}{2D(\bar\rho)} P_2(x,0) +\left(\frac{\sigma'(\bar\rho)}{2\sigma(\bar\rho)}-\frac{D'(\bar\rho)}{2D(\bar\rho)}\right)Q_1(x,0)^{2}
  \:.
\end{equation}
Setting
\begin{multline}
  \label{eq:DefP2t_Arb}
  P_2(x,t) = \frac{\sigma'(\bar\rho)}{16D(\bar\rho)}\erfc \left( \frac{x + \nu \sigma'(\bar\rho))}{2 \sqrt{D(\bar\rho)(1-t})} \right)
  \erfc \left(- \frac{x + \nu \sigma'(\bar\rho)}{2 \sqrt{D(\bar\rho)(1-t)}} \right) \\+\frac{\nu \sigma(\bar\rho)\sigma''(\bar\rho)}{8D(\bar\rho)^{2}}\left(\sqrt{\frac{D(\bar\rho)(1-t)}{\pi}} \e^{-\frac{(x + \nu \sigma'(\bar\rho))^2}{4D(\bar\rho)(1-t)} }+\frac{x+\nu\sigma'(\bar\rho)}{2}\erfc \left(- \frac{x + \nu \sigma'(\bar\rho)}{2 \sqrt{D(\bar\rho)(1-t)}} \right)\right)\erfc \left(\frac{x + \nu \sigma'(\bar\rho)}{2 \sqrt{D(\bar\rho)(1-t)}} \right) 
  + \tilde{P}_2(x,t)
  \:,
\end{multline}
we find that $\tilde{P}_2$ obeys
\begin{multline}
  \label{eq:P2source_Arb}
  \partial_t \tilde{P}_2 + D(\rb )\partial_x^2 \tilde{P}_2
  =\frac{\nu\sigma(\bar\rho)\sigma''(\bar\rho)}{8D(\bar\rho)}
  \frac{\e^{- \frac{(x + \nu \sigma'(\bar\rho))^2}{4D(\bar\rho)(1-t)}}}{\sqrt{D(\bar\rho)\pi(1-t)}}
  \erfc \left( - \frac{x}{2 \sqrt{D(\bar\rho)t}} \right) \\\hspace{1.5cm} +
  \frac{\e^{- \frac{(x
  + \nu \sigma'(\bar\rho))^2}{4D(\bar\rho)(1-t)}}D'(\bar\rho)\sigma(\bar\rho)(x+\nu\sigma'(\bar\rho))}{16D(\bar\rho)^{5/2}\sqrt{\pi}(1-t)^{3/2}}\left(\erfc \left(  \frac{x}{2 \sqrt{D(\bar\rho)t}}\right)-\erfc \left(  \frac{x+\nu\sigma'(\bar\rho)}{2 \sqrt{D(\bar\rho)(1-t)}}\right)\right)
  \:,
  \quad
  \tilde{P}_2(x,1) = 0
  \:.
\end{multline}
Then setting one more time:
\begin{equation}
    \tilde{P}_2(2\sqrt{D(\rb)}x,t)=\frac{\sigma (\rb ) D'(\rb )}{D(\rb )^{5/2}}
     \tilde{P}_{2d}(x,t)
    +\frac{\nu  \sigma (\rb ) \sigma ''(\rb )}{D(\rb )^{3/2}}
     \tilde{P}_{2s}(x,t)
    +\frac{\sigma (\rb ) D'(\rb )
  }{D(\rb)^2}
   \tilde{f}_{2p}\left(\frac{\frac{\nu  \sigma '(\rb )}{2 \sqrt{D(\rb )}}+x}{\sqrt{1-t}}\right),
\end{equation}
where we have defined
\begin{equation}
     \tilde{f}_{2p}(x)=
     \frac{x \: e^{-x^2}}{8 \sqrt{\pi} }  \left(\erfc\left(x\right)-2\right)
     -
     \frac{\e^{-2x^2}}{8 \pi}
     .
\end{equation}

\bigskip
Inserting that expression of $\tilde{P}_{2}$ in \eqref{eq:P2source_Arb} we find that $\tilde{P}_{2d}$ and $\tilde{P}_{2s}$ obey the following anti-diffusion equations:
\begin{align}
   \label{eq:P2tdsource_Arb}
  &\partial_t \tilde{P}_{2d} + \frac{1}{4}\partial_x^2 \tilde{P}_{2d}
  = \frac{\erfc\left(-\frac{x}{\sqrt{t}}\right) \left(2 x \sqrt{D(\rb )}+\nu \sigma
   '(\rb )\right) \exp \left(-\frac{\left(2 x \sqrt{D(\rb )}+\nu  \sigma '(\rb )\right)^2}{4 (1-t)
   D(\rb )}\right)}{16 \sqrt{\pi } (1-t)^{3/2}},\\
   & \partial_t \tilde{P}_{2s} + \frac{1}{4}\partial_x^2 \tilde{P}_{2s}
  = \frac{\erfc\left(-\frac{x}{\sqrt{t}}\right) \exp \left(-\frac{\left(2 x \sqrt{D(\rb )}+\nu 
   \sigma '(\rb )\right)^2}{4 (1-t) D(\rb )}\right)}{8 \sqrt{\pi(1 - t)}}.
    \label{eq:P2tssource_Arb}
\end{align}
We can implement the initial condition \eqref{boundary_order2_Arb} for $Q_2$ by defining  $\tilde{Q}_2$ as
\begin{multline}
  \label{eq:DefQ2t_Arb}
  Q_2(x,t) = \frac{\sigma(\rb)}{2D(\rb)} P_2(x,t)+ \left(\frac{\sigma'(\rb)}{2 \sigma(\rb)}-\frac{D'(\rb)}{2 D(\rb)}\right) Q_1(x,t)^2 \\
  + \frac{\sigma(\rb)\sigma'(\rb)}{32D(\rb)2}\erfc \left(  \frac{x}{2 \sqrt{D(\rb)t}}\right)\erfc \left(  \frac{-x}{2 \sqrt{D(\rb)t}}\right) + \tilde{Q}_2(\frac{x}{2\sqrt{D(\rb)}},t)
  \:.
\end{multline}
with $\tilde{Q}_2(x,0) = 0$.

\bigskip

As for $\tilde{P_2}$ we set: 
\begin{equation}
\label{eq:Q2t}
   \tilde{Q}_2(x,t) = \frac{\sigma (\rb )^2 D'(\rb ) \left(\tilde{Q}_{2d}(x,t)+\frac{1}{2}
   \tilde{f}_{2p}\left(\frac{x}{\sqrt{t}}\right)\right)}{D(\rb )^3}+\frac{\nu  \sigma (\rb )^2 \sigma''(\rb ) \left( \tilde{Q}_{2s}(x,t)+\sqrt{t} \tilde{f}_{2s}
   \left(\frac{x}{\sqrt{t}}\right)\right)}{D(\rb )^{5/2}},
\end{equation}
where we have introduced:
\begin{equation}
\tilde{f}_{2s}(x,t)=\frac{1}{16} \left(x (\erfc(x)-2)-\frac{e^{-x^2}}{\sqrt{\pi }}\right) \erfc(x).
\end{equation}
Injecting that expression in \eqref{eq:Q2t}, one deduces the following diffusion equations:
\begin{align}
  \partial_t \tilde{Q}_{2d} - \frac{1}{4}\partial_x^2 \tilde{Q}_{2d}
  &= \frac{x e^{-\frac{x^2}{t}} \erfc\left(-\frac{2 x \sqrt{D(\rb )}+\nu  \sigma '(\rb )}{2 \sqrt{(1-t) D(\rb )}}\right)}{16 \sqrt{\pi } t^{3/2}},
  &\tilde{Q}_{2d}(x,0) &=0 \:,
  \label{eq:Q2tdsource_Arb}\\
  \partial_t \tilde{Q}_{2s} - \frac{1}{4}\partial_x^2 \tilde{Q}_{2s}
  &= \frac{e^{-\frac{x^2}{t}} \erfc\left(-\frac{2 x \sqrt{D(\rb )}+\nu  \sigma '(\rb )}{2 \sqrt{(1-t) D(\rb )}}\right)}{16 \sqrt{\pi } \sqrt{t}}\:,
  &\tilde{Q}_{2s}(x,0) &=0 \:.\label{eq:Q2tssource_Arb}
\end{align}
The solutions of the equations for $\tilde{P}_{2d},\tilde{P}_{2s},\tilde{Q}_{2d}$
and $\tilde{Q}_{2s}$~(\ref{eq:P2tdsource_Arb},\ref{eq:P2tssource_Arb},\ref{eq:Q2tdsource_Arb},\ref{eq:Q2tssource_Arb}) are related by
\begin{align}
\tilde{P}_{2d}(x,t; \nu)
&=2 \sqrt{D(\rb )} \tilde{Q}_{2d}\left(\frac{\nu  \sigma '(\rb )}{2 \sqrt{D(\rb )}}+x,1-t; -\nu \right)
\label{P2dfromQ2d}
\\
\tilde{P}_{2s}(x,t;\nu)
&=-2 \tilde{Q}_{2s}\left(x-\frac{\nu  \sigma '(\rb )}{2 \sqrt{D(\rb )}},1-t;-\nu \right)
\label{P2sfromQ2s}
\end{align}
where we have intraduced a third argument $\pm \nu$ to notice that in the expression of $\tilde{Q}_{2d}$ and $\tilde{Q}_{2s}$ one has to replace $\nu$ by $-\nu$ to deduce $\tilde{P}_{2d}$ and $\tilde{P}_{2s}$. We now turn to the determination of these functions.

\subsubsection{The term in \texorpdfstring{$\sigma''$}{sigma}}

Our goal in this section is to determine the function $\tilde{Q}_{2s}$. Inserting the change of variable $\nu \to \frac{2 Y \sqrt{D(\rb )}}{\sigma '(\rb )}$ in~\eqref{eq:Q2tssource_Arb}, one gets:
\begin{equation}
    \partial_t \tilde{Q}_{2s} - \frac{1}{4}\partial_x^2 \tilde{Q}_{2s}
  = \frac{1}{16} \frac{e^{-\frac{x^2}{t}} \erfc\left(-\frac{x+Y}{\sqrt{1-t}}\right)}{\sqrt{\pi } \sqrt{t}}
\end{equation}
Formally we can write the solution as
\begin{equation}
  \tilde{Q}_{2s}(x,t|Y) = \int_0^t \dd t'\int_{-\infty}^\infty \dd y
  K(x-y | t-t') \frac{\e^{-y^2/t'}}{16\sqrt{\pi t'}}
  \erfc \left( - \frac{y+Y}{\sqrt{1-t'}} \right)
  \:,
\end{equation}
with
\begin{equation}
\label{heat_kernel}
  K(x|t) =
  \frac{\e^{-x^2/t}}{\sqrt{\pi t}}
\end{equation}
the heat kernel. The integral over $y$ can be performed
analytically~\cite{Owen:1980SM}. Since we only need $\tilde{P}_2$ at $t=0$
and $\tilde{Q}_2$ at $t=1$, we only compute
\begin{equation}
    \label{eq:DefIxY}
  \tilde{Q}_{2s}(x,1|Y) = \frac{\e^{-x^2}}{16 \sqrt{\pi}} \int_0^1 \dd t
  \erfc \left(-\frac{Y+t x}{\sqrt{1-t^2}} \right)
  \equiv I(x,Y)
  \:.
\end{equation}
To determine this function, we notice that the integral can be simplified by computing
\begin{equation}
    \partial_x I(x,Y) +2 x I(x,Y)
    = \int_{0}^1 t \frac{\e^{-\frac{x^2 + 2t x Y + Y^2}{1-t^2}}}{8 \pi\sqrt{1-t^2}} \dd t
    \:.
\end{equation}
Using the method of Refs.~\cite{Grabsch:2022SM,Grabsch:2023SM} which relies on a numerical evaluation of the integral, we find that
\begin{multline}
    \partial_x I(x) +2xI(x)= \frac{1}{8 \pi}
    \left[ x \: \erfc( x \sg{x+Y} ) - \sg{x+Y} \frac{\e^{-x^2}}{\sqrt{\pi}} \right]
    \\
    \times
    \left[ Y \: \erfc( Y \sg{x+Y} ) - \sg{x+Y} \frac{\e^{-Y^2}}{\sqrt{\pi}} \right]
    \:.
\end{multline}
Integrating this differential equation, we find that
\begin{equation}
   \label{Q2sT1}
    \tilde{Q}_{2s}(x,1) = I(x,Y)
    =
    \Theta(x+Y)\frac{\e^{-x^2}}{8 \sqrt{\pi}}
    + \frac{1}{16} \erfc(y \sg{x+Y})
    \left[
        Y \: \erfc(Y \sg{x+Y})
        - \sg{x+Y} \frac{\e^{-Y^2}}{\sqrt{\pi}}
    \right]
    \:.
\end{equation}
Using the relation~\eqref{P2dfromQ2d} we obtain immediately the expression of $\tilde{P}_{2s}$ at $t=0$, which will be useful below to compute $q(x,0)$,
\begin{equation}
    \label{P2tsT0}
    \tilde{P}_{2s}(x,0)=
    - \frac{\e^{-(x+Y)^2}}{4\sqrt{\pi}} \Theta(x)
    + \frac{1}{8} \erfc\left((x+Y)\sg{x} \right)
    \left[
    Y \: \erfc(-Y \sg{x}) + \sg{x}\frac{\e^{-Y^2}}{\sqrt{\pi}} 
    \right]
    \:.
\end{equation}

\subsubsection{The term in \texorpdfstring{$D'$}{D}}

We follow the same procedure to write $\tilde{Q}_{2d}$ from~\eqref{eq:Q2tdsource_Arb} as
\begin{equation}
    \tilde{Q}_{2d}(x,1) = \int_0^1 \dd t'\int_{-\infty}^\infty \dd y
  K(x-y | t-t') 
  \frac{y \: e^{-\frac{y^2}{t'}} \erfc\left(-\frac{y + Y }{ \sqrt{1-t'}}\right)}{16 \sqrt{\pi } (t')^{3/2}}
  \:.
\end{equation}
Performing the integral over $y$ using~\cite{Owen:1980SM}, we get,
\begin{equation}
    \tilde{Q}_{2d}(x,1) = x \: \e^{-x^2} \int_0^1 \frac{\dd t}{16 \sqrt{\pi}}
    \erfc \left( - \frac{t x + Y}{\sqrt{1-t^2}} \right)
    + \e^{-x^2} \int_0^1 \frac{\dd t}{16 \pi} \sqrt{\frac{1-t}{1+t}}
    \e^{-\frac{(t x + Y)^2}{1-t^2}}
    \:.
\end{equation}
We notice that this function can be expressed in terms of the integral $I(x,Y)$~\eqref{eq:DefIxY} computed previously as
\begin{equation}
    \tilde{Q}_{2d}(x,1) = \frac{1}{2} 
    \partial_x \left( I(Y,x) - I(x,Y) \right)
    \:.
\end{equation}
Therefore, we directly obtain from~\eqref{Q2sT1},
\begin{equation}
    \label{Q2dT1}
    \tilde{Q}_{2d}(x,1)=
    - \frac{\e^{-x^2 - Y^2}}{16 \pi}
    + \frac{x \: \e^{-x^2}}{8 \sqrt{\pi}} \Theta(x+Y)
    + \frac{1}{32} \erfc (Y \sg{x+Y})
    \left[ 
    \erfc (x \sg{x+Y}) + 2 Y \sg{x+Y} \frac{\e^{-x^2}}{\sqrt{\pi}}
    \right]
    \:.
\end{equation}
Using~\eqref{P2dfromQ2d}, we deduce
\begin{multline}
 \label{P2tdT0}
    \frac{\tilde{P}_{2d}(x,0)}{2 \sqrt{D(\rb)}}
    =
    - \frac{\e^{-(x+Y)^2 - Y^2}}{16 \pi}
    + \frac{(x+Y) \: \e^{-(x+Y)^2}}{8 \sqrt{\pi}} \Theta(x)
    \\
    + \frac{1}{32} \erfc (Y \sg{x})
    \left[ 
    \erfc ((x+Y) \sg{x}) + 2 Y \sg{x} \frac{\e^{-(x+Y)^2}}{\sqrt{\pi}}
    \right]
    \:.
\end{multline}

\subsubsection{Summary}

Using the relations (\ref{eq:DefQ2t_Arb},\ref{eq:Q2t},\ref{Q2sT1},\ref{P2tsT0}) and~(\ref{eq:ChangeFrame}), we obtain the solution of the MFT equations at order $2$ in $\lambda$, at initial and final time,
\begin{multline}
    \label{q2T1}
    q_2(x,1)=\frac{\sigma (\rb )}{32 D(\rb )^3} \Biggl(D(\rb ) \sigma '(\rb ) \erfc\left(\frac{x-\nu  \sigma '(\rb )}{2 \sqrt{D(\rb )}}\right)
   \erfc\left(\frac{\nu  \sigma '(\rb )-x}{2 \sqrt{D(\rb )}}\right)  -\left(\sigma (\rb ) d'(\rb
   )-D(\rb ) \sigma '(\rb )\right) \left(\erfc\left(\frac{\nu  \sigma '(\rb )-x}{2 \sqrt{D(\rb
   )}}\right)-2 \Theta (x)\right)^2 \\ 
   +\frac{\sigma (\rb ) d'(\rb ) e^{-\frac{3 \nu ^2 \sigma '(\rb )^2+2 x^2}{4 D(\rb )}}}{\pi 
   \sqrt{D(\rb )}} \Biggl(\sqrt{\pi
   } \e^{\frac{2 \nu ^2 \sigma '(\rb )^2+x^2+2 \nu  x \sigma '(\rb )}{4 D(\rb )}}
   \Bigg(\nu  \sigma '(\rb ) \Biggl(\sg{x} \erfc\left(\frac{\nu  \sg{x} \sigma
   '(\rb )}{2 \sqrt{D(\rb )}}\right) \\ -\erfc\left(\frac{x-\nu  \sigma '(\rb )}{2 \sqrt{D(\rb
   )}}\right)+2\Biggr) +x \left(\erfc\left(\frac{x-\nu  \sigma '(\rb )}{2 \sqrt{D(\rb
   )}}\right)-2\right)+2 \Theta (x) \left(x-\nu  \sigma '(\rb
   )\right)\Biggr)
   \\+\sqrt{D(\rb )} \left(\pi  e^{\frac{3 \nu ^2 \sigma '(\rb )^2+2 x^2}{4 D(\rb )}}
   \erfc\left(\frac{\nu  \sg{x} \sigma '(\rb )}{2 \sqrt{D(\rb )}}\right)
   \erfc\left(\frac{\sg{x} \left(x-\nu  \sigma '(\rb )\right)}{2 \sqrt{D(\rb )}}\right)-2
   \left(\e^{\frac{\nu  \sigma '(\rb ) \left(\nu  \sigma '(\rb )+4 x\right)}{4 D(\rb
   )}}+e^{\frac{\left(\nu  \sigma '(\rb )+x\right)^2}{4 D(\rb )}}\right)\right)\Biggr) \\
   -\frac{2 \nu  \sqrt{D(\rb )} \sigma (\rb ) \sigma ''(\rb ) e^{-\frac{\nu ^2 \sigma '(\rb )^2+x^2}{4
   D(\rb )}} \left(\sg{x} e^{\frac{x^2}{4 D(\rb )}} \erfc\left(\frac{\sg{x}
   \left(x-\nu  \sigma '(\rb )\right)}{2 \sqrt{D(\rb )}}\right)+e^{\frac{\nu  x \sigma '(\rb )}{2
   D(\rb )}} \erfc\left(\frac{x-\nu  \sigma '(\rb )}{2 \sqrt{D(\rb )}}\right)-2 \Theta
   (x) e^{\frac{\nu  x \sigma '(\rb )}{2 D(\rb
   )}}\right)}{\sqrt{\pi }} \\
   +\nu  \sigma (\rb ) \sigma ''(\rb ) \Bigg(\nu  \sigma '(\rb ) \left(\erfc\left(\frac{\nu 
   \sg{x} \sigma '(\rb )}{2 \sqrt{D(\rb )}}\right) \erfc\left(\frac{\sg{x}
   \left(x-\nu  \sigma '(\rb )\right)}{2 \sqrt{D(\rb )}}\right)-\erfc\left(\frac{x-\nu  \sigma
   '(\rb )}{2 \sqrt{D(\rb )}}\right)^2+2 \erfc\left(\frac{x-\nu  \sigma '(\rb )}{2 \sqrt{D(\rb
   )}}\right)\right)+ \\ 
   x \left(\erfc\left(\frac{x-\nu  \sigma '(\rb )}{2 \sqrt{D(\rb
   )}}\right)-2\right) \erfc\left(\frac{x-\nu  \sigma '(\rb )}{2 \sqrt{D(\rb )}}\right)\Bigg)\Biggr)
   \:,
\end{multline}
\begin{multline}
\label{q2T0}
    q_2(x,0)=\frac{\sigma (\rb )}{32 D(\rb
   )^3} \Bigg(D(\rb ) \sigma '(\rb ) \erfc\left(-\frac{\nu  \sigma '(\rb )+x}{2 \sqrt{D(\rb )}}\right)
   \erfc\left(\frac{\nu  \sigma '(\rb )+x}{2 \sqrt{D(\rb )}}\right)-\frac{2 \sigma (\rb )
    D'(\rb ) e^{-\frac{\left(\nu  \sigma '(\rb )+x\right)^2}{2 D(\rb )}}}{\pi } \\ 
   + \frac{\sigma (\rb ) D'(\rb ) \left(\nu  \sigma '(\rb )+x\right) e^{-\frac{\left(\nu  \sigma '(\rb
   )+x\right)^2}{4 D(\rb )}} \left(\erfc\left(\frac{\nu  \sigma '(\rb )+x}{2 \sqrt{D(\rb
   )}}\right)-2\right)}{\sqrt{\pi } \sqrt{D(\rb )}}
   - \left(\sigma (\rb ) D'(\rb )-D(\rb ) \sigma '(\rb )\right) \left(\erfc\left(-\frac{\nu 
   \sigma '(\rb )+x}{2 \sqrt{D(\rb )}}\right)-2 \Theta (x)\right)^2 \\
   + \frac{\sigma (\rb ) D'(\rb )\e^{-\frac{2 \nu ^2 \sigma '(\rb )^2+x^2+2 \nu  x \sigma '(\rb
   )}{4 D(\rb )}}}{\pi  \sqrt{D(\rb )}}\Bigg(\sqrt{\pi } e^{\frac{\nu ^2 \sigma '(\rb )^2}{4 D(\rb )}} \left(2 \Theta (x) \left(\nu  \sigma '(\rb )+x\right)-\nu  \sg{x} \sigma '(\rb )
   \erfc\left(-\frac{\nu  \sg{x} \sigma '(\rb )}{2 \sqrt{D(\rb )}}\right)\right) \\+ \sqrt{D(\rb )} \left(\pi  \e^{\frac{2 \nu ^2 \sigma '(\rb )^2+x^2+2 \nu  x \sigma '(\rb )}{4
   D(\rb )}} \erfc\left(-\frac{\nu  \sg{x} \sigma '(\rb )}{2 \sqrt{D(\rb
   )}}\right) \erfc\left(\frac{\sg{x} \left(\nu  \sigma '(\rb )+x\right)}{2 \sqrt{D(\rb
   )}}\right)-2\right) \\+ \frac{2 \nu  \sqrt{D(\rb )} \sigma (\rb ) \sigma ''(\rb ) e^{-\frac{\left(\nu  \sigma '(\rb
   )+x\right)^2}{4 D(\rb )}} \erfc\left(\frac{\nu  \sigma '(\rb )+x}{2 \sqrt{D(\rb
   )}}\right)}{\sqrt{\pi }}+\nu  \sigma (\rb ) \sigma ''(\rb ) \left(\nu  \sigma '(\rb )+x\right)
   \erfc\left(-\frac{\nu  \sigma '(\rb )+x}{2 \sqrt{D(\rb )}}\right) \erfc\left(\frac{\nu 
   \sigma '(\rb )+x}{2 \sqrt{D(\rb )}}\right) \\ 
   -\frac{2 \nu  \sqrt{D(\rb )} \sigma (\rb ) \sigma ''(\rb )}{\sqrt{\pi }} \Bigg(\frac{1}{2}
   \erfc\left(\frac{\sg{x} \left(\nu  \sigma '(\rb )+x\right)}{2 \sqrt{D(\rb )}}\right)
   \left(-\frac{\sqrt{\pi } \nu  \sigma '(\rb ) \erfc\left(-\frac{\nu  \sg{x} \sigma
   '(\rb )}{2 \sqrt{D(\rb )}}\right)}{\sqrt{D(\rb )}}-2 \sg{x} e^{-\frac{\nu ^2 \sigma '(\rb
   )^2}{4 D(\rb )}}\right) \\+2 \Theta (x) e^{-\frac{\left(\nu 
   \sigma '(\rb )+x\right)^2}{4 D(\rb )}}\Bigg)\Bigg)
   \:.
\end{multline}

\section{Cumulants and profiles for the integrated current}

\subsection{Cumulants}

We can directly compute the cumulants of $Q_T$ from the profiles at the first orders~(\ref{q1T1},\ref{q1T0},\ref{q2T1},\ref{q2T0}) using the relation~\eqref{eq:dpsidlambda}, which expanded in powers of $\lambda$ reads
\begin{equation}
  \label{eq:CumulPsi}
  \dt{\hat{\psi}}{\lambda} = \sum_{n=0}^\infty \frac{\lambda^n}{n!} \hat{\kappa}_{n+1}
  = \nu \sigma(\bar\rho)
  + \sum_{n=1}^\infty \int_0^\infty \left[ q_n(x,1) - q_n(x,0) \right] \dd x
  \:.
\end{equation}
Note that the first cumulant is already determined by the first term of~\eqref{eq:CumulPsi}, while the higher-order cumulants are encoded in the second. We thus get 
\begin{equation}
\label{eq:Kappa1SM}
    \hat{\kappa}_1 = \nu\sigma(\rb)
    \:,
\end{equation}
\begin{equation}
\label{eq:Kappa2SM}
    \hat{\kappa_2} = \int_0^\infty (q_1(x,1) - q_1(x,0)) \dd x
    =  \frac{\sigma(\rb)}{2\sqrt{D(\rb)}} \left(
      \frac{2}{\sqrt{\pi}} \e^{-\frac{(\nu \sigma'(\rb))^2}{4D(\rb)}}
      + \frac{\nu \sigma'(\rb)}{\sqrt{D(\rb)}} \erf \left( \frac{\nu \sigma'(\rb)}{2\sqrt{D(\rb)}} \right)
    \right)
    \:,
\end{equation}
\begin{multline}
\label{eq:Kappa3SM}
  \hat\kappa_3 =
  \frac{ \nu \sigma(\rb)}{16D(\rb)^{3}}\Bigg( 3\nu^{2}\sigma(\rb)\sigma'(\rb)^{2}\sigma''(\rb)+4D(\rb)\left(\sigma'(\rb)^{2}+\left(1-\frac{3\e^{-\frac{\nu^{2}\sigma'(\rb)^{2}}{2D(\rb)}}}{\pi}\right)\sigma(\rb)\sigma''(\rb)\right)\\+3\nu\sigma(\rb)\sigma'(\rb)\sigma''(\rb)\erf\left(\frac{\nu \sigma'(\rb)}{2\sqrt{D(\rb)}}\right)\left(-\frac{4\sqrt{D(\rb)}\e^{-\frac{\nu^{2}\sigma'(\rb)^{2}}{4D(\rb)}}}{\sqrt{\pi}} -\nu\sigma'(\rb) \erf\left(\frac{\nu \sigma'(\rb)}{2\sqrt{D(\rb)}}\right) \right)\Bigg) \\
  + \frac{D'(\rb)\sigma(\rb)^{2}}{8D(\rb)^{3}}\left(\nu\sigma'(\rb) -\frac{6\sqrt{D(\rb)} \erf\left(\frac{\nu \sigma'(\rb)}{2\sqrt{D(\rb)}}\right)}{\sqrt{\pi}}-3\nu\sigma'(\rb)\erf\left(\frac{\nu \sigma'(\rb)}{2\sqrt{D(\rb)}}\right)^{2}\right)
  \:,
\end{multline}
which correspond to Eqs.~(10,11) in the main text.

\subsection{Correlation profile}

In order to get the correlation profiles at the microscopic large time $T$ we remark that we can write their generating function as~\cite{Poncet:2021SM},
\begin{align}
\nonumber
\Phi(x) \equiv
    \frac{\moy{\rho(x,1)\e^{\lambda Q_T}}}{\moy{\e^{\lambda Q_T}}} &= \frac{\int \D \rho \D H\rho(x,1)
  \: \e^{-\sqrt{T} \left( S[\rho,H] + F[\rho(x,0)] - \lambda \hat{Q}[\rho] \right)}
 }
{\int \D \rho \D H
  \: \e^{-\sqrt{T} \left( S[\rho,H] + F[\rho(x,0)] - \lambda \hat{Q}[\rho] \right)}} \\
  \nonumber
  &\simeq  \frac{q(x,1)
  \: \e^{-\sqrt{T} \left( S[q,p] + F[q(x,0)] - \lambda \hat{Q}[q] \right)}
 }
{\e^{-\sqrt{T} \left( S[q,p] + F[q(x,0)] - \lambda \hat{Q}[q] \right)}} \\
&=q(x,1)
\label{correlation_saddle}
\end{align}
because $(q,p)$ is the saddle point of $ S[\rho,H] + F[\rho(x,0)] - \lambda \hat{Q}[\rho]$, using the expansion \eqref{eq:DefExppq} and the results of our computations for the profiles~(\ref{q1T1},\ref{q2T1}), we deduce
\begin{align}
    \label{eq:Phi1QTSM}
    \moy{\rho(x,1)Q_T}_c
    \simeq q_1(x,1) &= \sg{x} \frac{\sigma(\bar\rho)}{4D(\bar\rho)} \erfc
      \left(
      \sg{x} \frac{x - \nu \sigma'(\bar{\rho})}{2\sqrt{D(\bar\rho)}}
      \right) \:, \\
     \moy{\rho(x,1)Q_{T}^2}_c &\simeq q_2(x,1)
\end{align}
where $q_2$ is given by~\eqref{q2T1} and the index $c$ refers to the connected moments (for example $\moy{\rho(x,1)Q_T}_c=\moy{\rho(x,1)Q_T}-\moy{\rho(x,1)}\moy{Q_T}$ is the covariance, and higher order connected moments generalize it).

\section{Cumulants and profiles for the tracer}

\subsection{Cumulants}

Using the duality mapping presented in \ref{subsection:duality}, we deduce the first three normalized cumulants of a tracer initially at the origin:
\begin{equation}
\label{eq:Kappa1tracerSM}
  \frac{\moy{X_T}}{\sqrt{T}} 
  \underset{T \to \infty}{\simeq}
  \nu \frac{\sigma(\rb)}{\rb}
  \:,
\end{equation}

\begin{equation}
    \label{eq:Kappa2tracerSM}
    \frac{\moy{X_T^2}_c}{\sqrt{T}} 
  \underset{T \to \infty}{\simeq}
  \frac{\sigma (\rb )}{2 \rb ^3 D(\rb )} \left(\nu  \left(\sigma (\rb )-\rb  \sigma '(\rb )\right)
   \erf\left(\frac{\nu  \left(\sigma (\rb )-\rb  \sigma '(\rb )\right)}{2 \sqrt{\rb ^2 D(\rb
   )}}\right)+\frac{2 \sqrt{\rb ^2 D(\rb )} \exp \left(-\frac{\nu ^2 \left(\sigma (\rb )-\rb 
   \sigma '(\rb )\right)^2}{4 \rb ^2 D(\rb )}\right)}{\sqrt{\pi }}\right)
\end{equation}
\begin{multline}
\label{eq:Kappa3tracerSM}
  \frac{\moy{X_T^3}_c}{\sqrt{T}} 
  \underset{T \to \infty}{\simeq}
\frac{\sigma (\rb )}{16 \rb ^5 D(\rb )^3} \Biggl(
\left(\rb  D'(\rb )+2 D(\rb )+\nu ^2 \left(\rb  \sigma '(\rb )-\sigma (\rb )\right) \sigma
   ''(\rb )\right)
   \\
   \times
\frac{12 \rb  \sqrt{D(\rb )} \sigma (\rb )}{\sqrt{\pi }} \erf\left(\frac{\nu 
   \left(\sigma (\rb )-\rb  \sigma '(\rb )\right)}{2 \rb  \sqrt{D(\rb )}}\right) \exp
   \left(-\frac{\nu ^2 \left(\sigma (\rb )-\rb  \sigma '(\rb )\right)^2}{4 \rb ^2 D(\rb )}\right) \\ 
   +\nu  \sigma (\rb ) \left(\sigma (\rb )-\rb  \sigma '(\rb )\right)
   \Biggl(2 \left(\rb  D'(\rb )+2 D(\rb )\right) \left(3 \erf\left(\frac{\nu  \left(\sigma
   (\rb )-\rb  \sigma '(\rb )\right)}{2 \rb  \sqrt{D(\rb )}}\right)^2-1\right)\\
   -3 \nu ^2\left(\sigma (\rb )-\rb  \sigma '(\rb )\right) \sigma ''(\rb ) \left(\erf\left(\frac{\nu 
   \left(\sigma (\rb )-\rb  \sigma '(\rb )\right)}{2 \rb  \sqrt{D(\rb
   )}}\right)^2-1\right)\Biggr) \\ 
   +4 \nu  D(\rb ) \left(\rb ^2 \sigma (\rb ) \sigma ''(\rb )
   \left(1-\frac{3 \exp \left(-\frac{\nu ^2 \left(\sigma (\rb )-\rb  \sigma '(\rb )\right)^2}{2 \rb
   ^2 D(\rb )}\right)}{\pi }\right)+\left(\sigma (\rb )-\rb  \sigma '(\rb
   )\right)^2\right) \Biggr)
  \:.
\end{multline}

\subsection{Correlation profiles}

The duality mapping~\eqref{eq:MappingDens} is written for the stochastic fields $\rho$ and $j$. In the limit $T \to \infty$, the MFT action is dominated by a specific realisation of these fields, and in particular an optimal density given by $q(x,t)$. Therefore, we can deduce the correlation profiles in the reference frame of the tracer from those obtained for $Q_T$~\eqref{correlation_saddle} as
\begin{equation}
    \Phi_T(x) \equiv \frac{\moy{\rho(x + X_T/\sqrt{T},1) \e^{\lambda X_T}}}{\moy{\e^{\lambda X_T}}}
    = \frac{1}{\Phi(y(x))} \Bigg|_{D \to \tilde{D}, \sigma \to \tilde\sigma, \nu \to \tilde\nu}
    \:,
    \quad
    y(x) = \int_0^x \Phi(x')\dd x'
    \:,
\end{equation}
where we have again used the scaling of Section~\ref{sec:MicroToMacro} to express the correlations in terms of the macroscopic density $\rho(x,1)$ which represents the microscopic system at the observation time $T$. For instance, we obtain at first order in $\lambda$,
\begin{equation}
    \moy{X_T \: \rho \left(x + \frac{X_T}{\sqrt{T}}, 1 \right)}_c
    \underset{T \to \infty}{\simeq}
    \frac{\sigma(\rb)}{4 \rb D(\rb)} \erfc \left(
    \frac{x - \nu \frac{\sigma(\rb)-\rb\sigma'(\rb)}{\rb}}{2 \sqrt{D(\bar\rho)}}
    \right)
    \:,
    \quad \text{for } x > 0
    \:.
\end{equation}
As announced in the main text, the spatial dependence of this profile is identical to the one obtained in the case of the current~\eqref{eq:Phi1QTSM}, but with $\sigma'(\rb)$ replaced by $\frac{\sigma(\rb)-\rb \sigma'(\rb)}{\rb}$.

\section{Large asymmetry: ballistic limit}

The situation of driven diffusive system can
be though as representing the dynamics of driven ballistic systems at a large but finite time $T$. Indeed, for
fixed $F$ and large $T$, choosing $\nu =F\sqrt{T}$ should allow to describe the properties of the system at time $T$. An interesting question is to look at the limit of large asymmetry \( \nu \to \infty \) at fixed $F$. Can we obtain the large times limit properties of the initial ballistic system by taking that limit? This convergence is not a priori obvious, as the hydrodynamic scalings in diffusive and ballistic systems are of different nature: for a system of size \( L \), the time scales like \( L^2 \) for diffusive dynamics, and like \( L \) for the ballistic case. Then there is no reason that the limits $\nu\to\infty$ and $T\to\infty$ commute. Nevertheless setting $\nu = F\sqrt{T}$ and keeping only the dominant term in
 \eqref{eq:Kappa1tracerSM}, \eqref{eq:Kappa2tracerSM} and \eqref{eq:Kappa3tracerSM}, we obtain for the tracer
\begin{align}
    \label{eq:MeanXtBallisticSM}
  \moy{ X_{T}}
    &\underset{T \to \infty}{\simeq} T \: F \: \frac{\sigma(\rb)}{\rb} \:, \\
    \label{eq:VarXtBallisticSM}
    \moy{ X^2_{T}}_c
    &\underset{T \to \infty}{\simeq} T \: \abs{F} \: \frac{\sigma(\rb)}{2 \rb^3 D(\rb)} 
    \abs{\sigma(\rb) - \rb \sigma'(\rb)} \:, \\
    \label{eq:Kappa3XtBallisticSM}
    \moy{ X^3_{T}}_c
    &\underset{T \to \infty}{\simeq} T \: F \: \frac{\sigma(\rb)}{4 \rb^3 D(\rb)^2} \bigg[
    \sigma''(\rb) \sigma(\rb)
    + \left( \sigma(\rb) - \rb \sigma'(\rb) \right) 
    \left( \frac{3 \sigma(\rb) - \rb \sigma'(\rb)}{\rb^2} + \frac{\sigma(\rb) D'(\rb)}{\rb D(\rb)} \right) \bigg].
\end{align}
These results correspond  to Eqs.~(15-17) of the main text.
Performing the same procedure for the integrated current given by equations~(\ref{eq:Kappa1SM}-\ref{eq:Kappa3SM}), we obtain
\begin{align}
    \moy{Q_T}=\sqrt{T}\hat{\kappa}_1  &\underset{T\to \infty}{\simeq} T \: \abs{F} \:\sigma(\rb)\\
    \moy{Q_T^{2}}_c=\sqrt{T}\hat{\kappa}_2  &\underset{T\to  \infty}{\simeq} T \: \abs{F} \:\frac{\left|\sigma (\rb ) \sigma '(\rb ) \right|}{2 D(\rb )}\\
     \moy{Q_T^{3}}_c=\sqrt{T}\hat{\kappa}_3  &\underset{T\to  \infty}{\simeq} T \: \abs{F} \:\frac{\sigma (\rb )}{4 D(\rb )^2} \bigg[
    \sigma(\rb) \left(\sigma''(\rb)-\frac{D'(\rb)\sigma'(\rb)}{D(\rb)}\right)
    + 
    \sigma'(\rb)^{2} \bigg].
\end{align}

\section{Extension: correlations between two tracers or two currents}

In this section, we demonstrate how our results on the correlation profiles can be used to derive, as a byproduct, the correlations between two integrated currents measured at different points. In turn, this allows to obtain the correlations between two tracers using the duality mapping~\eqref{eq:MappingCurrTracGen}. We thus consider the integrated current through a point $x$ up to time $T$, which reads in terms of the macroscopic fields,
\begin{equation}
    Q_T(x) \equiv \sqrt{T} \int_0^t j(x,t')\dd t'
    \:.
\end{equation}
Note that the macroscopic variable $x$ corresponds to a microscopic position $x \sqrt{T}$. 
The joint cumulant generating function of two integrated currents thus takes the form,
\begin{equation}
    \label{eq:DefJointMoments}
\moy{\e^{\lambda Q_T(0) + \chi Q_T(\xi)}} =
\int \D \rho \D H
\e^{-\sqrt{T}(S[\rho,H] + F[\rho(x,0)] - \lambda \hat{Q}_0[\rho] - \chi \hat{Q}_\xi[\rho])}
\:,
\end{equation}
where we have defined
\begin{equation}
    \label{eq:DefHatQxi}
    \hat{Q}_x[\rho] = \nu \sigma(\rb) + \int_{x}^\infty [ \rho(x',1) - \rho(x',0)] \dd x'
    \:,
\end{equation}
by analogy with the $x=0$ case~\eqref{eq:DefHatQ}. In the long time limit \( T \to \infty \), these integrals can again be evaluated by a saddle point method, which yields
\begin{equation}
\label{eq:PsiJoint}
\hat{\psi}_{\xi}(\lambda, \chi) = 
\lim_{T \to \infty} \frac{1}{\sqrt{T}} \ln \moy{\e^{\lambda Q_T(0) + \chi Q_T(\xi)}} 
= \lambda \hat{Q}_0[q] + \chi \hat{Q}_\xi[q] - S[q, p] - F[q(x, 0)].
\end{equation}
where we have denoted \( (q, p) \) the saddle point of \( (\rho, H) \), for given values of $\lambda$ and $\chi$. The key point is that, to obtain the covariance between the two currents at long time $T$, given by
\begin{equation}
    \label{eq:CovCurrents0}
\frac{\moy{Q_T(0) Q_T(\xi) }_c}{\sqrt{T}}\underset{T\to \infty}{\simeq} 
\frac{\partial^2 \hat{\psi}_{\xi}}{\partial \lambda \partial \chi}\bigg|_{\lambda=0,\chi=0}
\:,
\end{equation}
we only need the solution for $\chi=0$ computed in Section~\ref{sec:PertSol}. Indeed, since $(q,p)$ minimise the action in~\eqref{eq:DefJointMoments}, we have that
\begin{equation}
    \frac{\delta}{\delta \rho} \left(
    S[\rho,H] + F[\rho(x,0)] - \lambda \hat{Q}_0[\rho] - \chi \hat{Q}_\xi[\rho])
    \right) \Bigg|_{(q,p)}
    = 0
\end{equation}
and
\begin{equation}
    \frac{\delta}{\delta H} \left(
    S[\rho,H] + F[\rho(x,0)] - \lambda \hat{Q}_0[\rho] - \chi \hat{Q}_\xi[\rho])
    \right) \Bigg|_{(q,p)}
    = 0
    \:.
\end{equation}
Therefore, the derivative with respect to $\chi$ of~\eqref{eq:PsiJoint} is given by
\begin{equation}
    \frac{\dd }{\dd \chi} \hat{\psi}_{\xi}
    = \hat{Q}_\xi[q]
    \:.
\end{equation}
Using this into the expression of the covariance~\eqref{eq:CovCurrents0}, and using that $\hat{Q}_\xi[q]$~\eqref{eq:DefHatQxi} is linear in $q$, we obtain
\begin{equation}
\frac{\moy{Q_T(0) Q_T(\xi) }_c}{\sqrt{T}}\underset{T\to \infty}{\simeq} \frac{\partial \hat{Q}_\xi[q]}{\partial \lambda}\bigg|_{\lambda=0,\chi=0}
= \hat{Q}_\xi \left[ \dep{q}{\lambda} \bigg|_{\lambda=0,\chi=0} \right] 
= \hat{Q}_\xi \left[ \dep{}{\lambda} \left(q \Big|_{\chi=0} \right) \bigg|_{\lambda=0} \right] 
\:.
\end{equation}
This last relation involves only the optimal profile for $\chi = 0$, which is the one computed previously in Section~\ref{sec:PertSol}. Inserting the expression of $q$ at first order in $\lambda$~(\ref{q1T1},\ref{q1T0}), we directly obtain,
\begin{multline}
    \label{eq:current_covariance}
    \frac{\moy{Q_T(0) Q_T(\xi) }_c}{\sqrt{T}}\simeq 
    \frac{\sigma (\bar\rho )}{4 D(\bar\rho )}\Bigg(\frac{2 \sqrt{D(\bar\rho )} e^{-\frac{\left(\xi +\nu  \sigma '(\bar\rho )\right)^2}{4 D(\bar\rho )}}}{\sqrt{\pi }}-\left(\xi +\nu 
   \sigma '(\bar\rho )\right) \erfc\left(\frac{\xi +\nu  \sigma '(\bar\rho )}{2 \sqrt{D(\bar\rho
   )}}\right)\\
   +\left(\nu  \sigma '(\bar\rho )-\xi \right) \erfc\left(\frac{\xi -\nu  \sigma '(\bar\rho )}{2 \sqrt{D(\bar\rho
   )}}\right)+\frac{2 \sqrt{D(\bar\rho )} e^{-\frac{\left(\xi -\nu  \sigma '(\bar\rho )\right)^2}{4 D(\bar\rho )}}}{\sqrt{\pi }}\Bigg)
\end{multline}

From the result~\eqref{eq:current_covariance} and the duality mapping~\eqref{eq:MappingCurrTracGen}, we obtain the correlation at large time between two tracers $\tilde X_t$ and $\tilde Y_t$, initially placed at $\tilde{X}_0=0$ and $\tilde{Y}_0= \xi \sqrt{T}$ by performing the substitution
\begin{equation}
    D(\rho) \to \frac{1}{\rho^2} D \left( \frac{1}{\rho} \right)
    \:,
    \quad
    \sigma(\rho) \to \rho \: \sigma \left( \frac{1}{\rho} \right)
    \:,
    \quad
    \nu \to -\nu
    \:,
    \quad
    \rb \to \frac{1}{\rb}
    \:.
\end{equation}
Explicitly, this gives,
\begin{multline}
    \label{eq:tracers_covariance}
   \frac{\moy{\tilde{X}_T\tilde{Y}_T}_c}{\sqrt{T}}  =
\frac{\sigma (\rb )}{4 \rb ^3 D(\rb )} \Biggl(\xi  \left(\erf\left(\frac{\xi -\nu  \rb  \sigma '(\rb )+\nu  \sigma (\rb )}{2 \sqrt{\rb
   ^2 D(\rb )}}\right)+\erf\left(\frac{\xi +\nu  \rb  \sigma '(\rb )-\nu  \sigma (\rb )}{2
   \sqrt{\rb ^2 D(\rb )}}\right)-2\right) \\ +\nu  \left(\sigma (\rb )-\rb  \sigma '(\rb )\right)
   \left(\erf\left(\frac{\xi -\nu  \rb  \sigma '(\rb )+\nu  \sigma (\rb )}{2 \sqrt{\rb ^2
   D(\rb )}}\right)-\erf\left(\frac{\xi +\nu  \rb  \sigma '(\rb )-\nu  \sigma (\rb )}{2
   \sqrt{\rb ^2 D(\rb )}}\right)\right) \\ +\frac{2 \sqrt{\rb ^2 D(\rb )} \left(1+\e^{\frac{\xi  \nu 
   \left(\sigma (\rb )-\rb  \sigma '(\rb )\right)}{\rb ^2 D(\rb )}}\right) \exp
   \left(-\frac{\left(\xi -\nu  \rb  \sigma '(\rb )+\nu  \sigma (\rb )\right)^2}{4 \rb ^2 D(\rb
   )}\right)}{\sqrt{\pi }}\Biggr)
  \:.
\end{multline}
This corresponds to Eq.~(20) in the main text.

\section{The weakly asymmetric simple exclusion process (WASEP)}
\label{section:hydrodynamic_wasep}

In this section, we show how we can obtain exact macroscopic equations for the WASEP from the microscopic description of the asymmetric simple exclusion process (ASEP). These equations will be useful in Section~\ref{sec:Conjec} below to deduce a boundary relation for the cumulant generating function of the current and the density profiles. We also give the explicit form of the first cumulants of $Q_T$ and $X_T$ in the case of the WASEP, deduced from our general results. 

\subsection{Generalities on the ASEP}

We consider an infinite one-dimensional lattice and introduce the occupation numbers \( \eta_i = 0 \) or \( 1 \) for site \( i \). Particles can jump to the left with rate $p_-$ and to the right with rate $p_+$, only if the neighbouring site is empty. The master equation associated with the probability \( P_t(\{ \eta \}) \), where \( \{ \eta \} \equiv \{ \eta_i \}_{i \in \mathbb{Z}} \) is a given configuration of the system, reads
\begin{equation}
  \label{eq:MasterEqASEP}
  \partial_t P_t(\{ \eta \}) =
  \sum_i 
  \left[c_i( \{ \eta \}_i^+ ) P_t( \{ \eta \}_i^+) - c_i( \{ \eta \} ) P_t( \{ \eta \}) \right]
  \:,
  \quad
  c_i( \{ \eta \} )
  = p_+ \: \eta_i (1-\eta_{i+1})+ p_- \:  \eta_{i+1} (1-\eta_{i})
  \:.
\end{equation}
where $\{ \eta \}_i^+$ is the configuration in which $\eta_i$ and
$\eta_{i+1}$ have been exchanged. The stationary distribution is given by the product of independent Bernoulli random variables for each site (the same as for the SEP),
\begin{equation}
  \label{eq:ProbWASEPstat}
  P_{\mathrm{stat}}(\{ \eta \})
  = \prod_i P_{\mathrm{Bern}}(\eta_i)
  \:,
  \quad
  P_{\mathrm{Bern}}(\eta_i = 1) = \bar\rho
  \:.
\end{equation}

\subsection{Microscopic equations for the ASEP}
\label{sec:ASEP}

From the master equation~(\ref{eq:MasterEqASEP}), we can write the time evolution of the generating function of the current through the origin $\Qt$, which counts the number of particles, that have jumped from site $0$ to site $1$, minus the number of particles that have jumped from site $1$ to $0$.
\begin{align}
  \partial_t \moy{\e^{\lambda \Qt}}
  =& \partial_t \sum_{\{ \eta \}} \e^{\lambda \Qt[\{ \eta \}]} P_t( \{ \eta \} )
    \nonumber
  \\
  = & \sum_{\{ \eta \}}  P_t( \{ \eta \} ) c_0( \{ \eta \} ) \e^{\lambda \Qt[\{\eta\}]}
    \left( \e^{\lambda \Qt[\{\eta\}_0^+]} -   \e^{\lambda \Qt[\{\eta\}]}  \right)
    \:.
\end{align}
Since the current is only changed when a particle jumps between sites 0 and 1, with the expression for $c_i$~(\ref{eq:MasterEqASEP}), we get

\begin{equation}
  \partial_t  \moy{\e^{\lambda \Qt}}
  = p_+ (\e^\lambda-1) \moy{\eta_0(1-\eta_1) \e^{\lambda \Qt}}
  + p_-(\e^{-\lambda}-1) \moy{\eta_1(1-\eta_0) \e^{\lambda \Qt}}
  \:.
\end{equation}
And then,
\begin{equation}
  \label{eq:EvolCGFmicro}
  \partial_t  \ln \moy{\e^{\lambda \Qt}}
  =
   p_+( \e^\lambda-1) \frac{\moy{\eta_0(1-\eta_1) \e^{\lambda \Qt}}}{\moy{\e^{\lambda \Qt}}}
  + p_-( \e^{-\lambda}-1) \frac{\moy{\eta_1(1-\eta_0) \e^{\lambda \Qt}}}{\moy{\e^{\lambda \Qt}}}
  \:.
\end{equation}
For $p_+=p_-$, we recover the one for the SEP~\cite{Grabsch:2022SM}.
Similarly, we can compute
\begin{multline}
  \partial_t \moy{\eta_1 \e^{\lambda \Qt}}
  = p_- \moy{\eta_2 (1-\eta_1) \e^{\lambda \Qt}}
  - p_+ \moy{\eta_1(1-\eta_2) \e^{\lambda \Qt}}
  \\
  + p_+ \e^\lambda \moy{\eta_0(1-\eta_1) \e^{\lambda \Qt}}
  - p_- \moy{\eta_1(1-\eta_0) \e^{\lambda \Qt}}
  \:.
\end{multline}
We can rewrite the second line of the above equation
using~(\ref{eq:EvolCGFmicro}), since
\begin{multline}
  p_+ \e^\lambda \moy{\eta_0(1-\eta_1) \e^{\lambda \Qt}}
  - p_- \moy{\eta_1(1-\eta_0) \e^{\lambda \Qt}}
  \\
  = - \frac{1}{\e^{-\lambda}-1} \left(
    p_+( \e^\lambda-1) \frac{\moy{\eta_0(1-\eta_1) \e^{\lambda \Qt}}}{\moy{\e^{\lambda \Qt}}}
  + p_-( \e^{-\lambda}-1) \frac{\moy{\eta_1(1-\eta_0) \e^{\lambda \Qt}}}{\moy{\e^{\lambda \Qt}}}
  \right)
  \:.
\end{multline}
We thus obtain
\begin{equation}
  \label{eq:EvolProf1Micro}
  \partial_t \frac{\moy{\eta_1 \e^{\lambda \Qt}}}{\moy{\e^{\lambda \Qt}}}
  =  p_- \frac{\moy{\eta_2 (1-\eta_1) \e^{\lambda \Qt}}}{\moy{\e^{\lambda \Qt}}}
  - p_+ \frac{\moy{\eta_1(1-\eta_2) \e^{\lambda \Qt}}}{\moy{\e^{\lambda \Qt}}}
  - \partial_t \ln \moy{\e^{\lambda \Qt}}
  \left( \frac{\moy{\eta_1 \e^{\lambda \Qt}}}{\moy{\e^{\lambda \Qt}}}
    + \frac{1}{\e^{-\lambda}-1}  \right)
  \:.
\end{equation}
Using the same procedure, we get
\begin{equation}
  \label{eq:EvolProf0Micro}
  \partial_t \frac{\moy{\eta_0 \e^{\lambda \Qt}}}{\moy{\e^{\lambda \Qt}}}
  =  p_+ \frac{\moy{\eta_{-1} (1-\eta_0) \e^{\lambda \Qt}}}{\moy{\e^{\lambda \Qt}}}
  - p_- \frac{\moy{\eta_0(1-\eta_{-1}) \e^{\lambda \Qt}}}{\moy{\e^{\lambda \Qt}}}
  - \partial_t \ln \moy{\e^{\lambda \Qt}}
  \left( \frac{\moy{\eta_0 \e^{\lambda \Qt}}}{\moy{\e^{\lambda \Qt}}}
    + \frac{1}{\e^{\lambda}-1}  \right)
  \:.
\end{equation}
We now use these equations to derive equations for the WASEP at the macroscopic scale.

\subsection{Microscopic equations for the WASEP}

We define the WASEP as follows. We choose an observation time $T>0$, and set the hopping rates of the ASEP as
\begin{equation}
  \label{eq:JumpRatesWeakAsym}
    p_+ = 1 + \frac{\nu}{\sqrt{T}}
    \:,
    \quad
    p_- = 1 - \frac{\nu}{\sqrt{T}}
    \:,
\end{equation}
with $\nu \in [-\sqrt{T}, \sqrt{T}]$. We then consider the integrated current $Q_T$ up to time $T$ or the displacement $X_T$ of a tracer at that same time in the ASEP with~\eqref{eq:JumpRatesWeakAsym}. Obtaining an equation for the time evolution of $\moy{\e^{\lambda Q_T}}$ in the WASEP is therefore more tricky than in the ASEP, because the parameters of the model~\eqref{eq:JumpRatesWeakAsym} change with the observation time $T$. To handle this, we introduce the different notations for the cumulant generating functions of the ASEP and the WASEP
\begin{equation}
    \label{eq:DefPsiWASEP}
    \psi_{\mathrm{ASEP}}(\lambda,t,p_+) \equiv \ln \moy{\e^{\lambda Q_t}}
    \:,
    \quad
    \psi_{\mathrm{WASEP}}(\lambda,T,\nu)
    = \psi_{\mathrm{ASEP}}\left( \lambda, T, 1+ \frac{\nu}{\sqrt{T}} \right)
    \:,
\end{equation}
where the averaging in the first equation is performed over the solution of the master equation~\eqref{eq:MasterEqASEP} for the ASEP with $p_+$ fixed. Note that we do not write the dependence on $p_-$ since we use the convention $p_+ + p_- = 2$.
Similarly, we define the correlation profiles in both cases as
\begin{equation}
    \label{eq:DefWsiWASEP}
    w_{\mathrm{ASEP}}(\lambda,t,r,p_+) \equiv 
    \frac{\moy{\eta_r(t) \: \e^{\lambda Q_t}}}{\moy{\e^{\lambda Q_t}}}
    \:,
    \quad
    w_{\mathrm{WASEP}}(\lambda,T,r,\nu) \equiv 
    w_{\mathrm{ASEP}} \left(\lambda,T,r, 1 +\frac{\nu}{\sqrt{T}} \right)
    \:.
\end{equation}
We can therefore deduce the equations satisfied in the WASEP from those obtained for the ASEP in Section~\ref{sec:ASEP}. For instance, taking the time derivative of~\eqref{eq:DefPsiWASEP}, we get
\begin{equation}
    \label{eq:EvolPsiWASEPfromASEP}
    \partial_T \psi_{\mathrm{WASEP}}
    = \left( \partial_t \psi_{\mathrm{ASEP}}
    - \frac{\nu}{2 T^{3/2}} \partial_{p_+} \psi_{\mathrm{ASEP}}
    \right) \Big|_{t=T,p_+ = 1 + \nu/\sqrt{T}}
    = - \frac{\nu}{2T} \partial_\nu \psi_{\mathrm{WASEP}}
    + \partial_t \psi_{\mathrm{ASEP}}\Big|_{t=T,p_+ = 1 + \nu/\sqrt{T}}
    \:,
\end{equation}
where we have used that $\partial_\nu \psi_{\mathrm{WASEP}} = \frac{1}{\sqrt{T}} \partial_{p_+} \psi_{\mathrm{ASEP}}$. Similarly, we obtain that
\begin{equation}
    \label{eq:EvolwWASEPfromASEP}
    \partial_T w_{\mathrm{WASEP}}
    =
    - \frac{\nu}{2T} \partial_\nu w_{\mathrm{WASEP}}
    +\partial_t w_{\mathrm{ASEP}} \Big|_{t=T,p_+ = 1 + \nu/\sqrt{T}}
    \:.
\end{equation}
Combined with~\eqref{eq:EvolCGFmicro} and~(\ref{eq:EvolProf1Micro},\ref{eq:EvolProf0Micro}),
these equations describe the time evolution of the cumulants and the correlation profiles in the WASEP. We now use these results to derive equations describing the WASEP at the macroscopic scale.

\subsection{Macroscopic equations for the WASEP}
\label{eq:WASEPmacro}

At the macroscopic scale, one can define the density and current fields for the WASEP by~\eqref{eq:DefMacroFieldsFromLattice} (with $n_i$ replaced by $\eta_i$). Since this model can be described within the MFT formalism, we have that the cumulant generating function is given by~\eqref{eq:PsiFromMFT}, i.e.,
\begin{equation}
    \label{eq:ScalingPsiWASEP}
    \psi_{\mathrm{WASEP}}(\lambda, T, \nu)
    \underset{T \to \infty}{\simeq} \sqrt{T} \: \hat{\psi}_{\mathrm{WASEP}}(\lambda,\nu)
    \:,
    \quad \text{with} \quad
    \dt{}{\lambda}\hat\psi_{\mathrm{WASEP}}
    = \int_0^\infty \left[ q(x,1) - q(x,0) \right] \dd x
    \:.
\end{equation}
Importantly, the solution $q$ of the MFT equation also describes the correlation profiles~\eqref{eq:DefCorrelMicro}, since from~\eqref{correlation_saddle},
\begin{equation}
    \label{eq:ScalingProfWASEP}
    w_{\mathrm{WASEP}}(\lambda,T,r,\nu)
    \underset{T \to \infty}{\simeq}
    \frac{\moy{ \rho(x,1) \e^{\lambda Q_T} }}{\moy{\e^{\lambda Q_T} }}
    = q(x,1) \equiv \Phi(x)
    \:,
    \quad
    \text{with}
    \quad
    x = \frac{r-\frac{1}{2}}{\sqrt{T}}
    \:,
\end{equation}
where we have introduced a shift $-\frac{1}{2}$ so that the site \( 0 \) corresponds to \( 0^- \) and the site \( 1 \) to \( 0^+ \). Inserting the scaling forms~(\ref{eq:ScalingPsiWASEP},\ref{eq:ScalingProfWASEP}) into the evolution equations~(\ref{eq:EvolPsiWASEPfromASEP},\ref{eq:EvolwWASEPfromASEP}) combined with~\eqref{eq:EvolCGFmicro} and~(\ref{eq:EvolProf1Micro},\ref{eq:EvolProf0Micro}), we obtain, at leading order in $T$, boundary conditions satified by $\Phi$,
\begin{equation}
    \label{continuity:q}
    (\e^\lambda - 1) \Phi(0^-) (1 - \Phi(0^+)) 
    + (\e^{-\lambda} - 1) \Phi(0^+) (1 - \Phi(0^-)) 
    = 0
    \:,
\end{equation}
\begin{equation}
    \label{eq:boundary_barpsi}
    \partial_x \Phi(0^\pm) -
    \left(\psi_{\mathrm{WASEP}}+\nu\partial_\nu\psi_{\mathrm{WASEP}}\right) 
    \frac{1}{2} \left( \Phi(0^\pm) + \frac{1}{e^{\mp\lambda} - 1} \right) 
    - 2\nu \Phi(0^\pm) (1 - \Phi(0^\pm)) = 0
    \:.
\end{equation}
Equivalently, these equations can be rewritten as
\begin{equation}
    \label{eq:lambda_wasep}
    \e^{\lambda}=\frac{(1-\Phi(0^-))\Phi(0^+)}{(1-\Phi(0^+))\Phi(0^-)}
    \:,
    \quad
    \frac{\partial_x \Phi(0^+)}{\Phi(0^+)(1-\Phi(0^+))}
    = \frac{\partial_x \Phi(0^-)}{\Phi(0^-)(1-\Phi(0^-))}
    \:,
\end{equation}
\begin{equation}
    \label{eq:psi_wasep}
    \psi_{\mathrm{WASEP}}+\nu\partial_\nu\psi_{\mathrm{WASEP}} = 
    -2\left(\frac{\partial_x \Phi(x)|_{0^+}}{2 \Phi(0^+) (1 - \Phi(0^+))}-2\nu\right)(\Phi(0^+) - \Phi(0^-))
    \:,
\end{equation}
In this form, these boundary conditions will be a basis to obtain the boundary conditions for any driven diffusive system in Section~\ref{sec:boundary_conditions}.

\subsection{Cumulants and correlations profiles}

Inserting the transport coefficients of the WASEP $D(\rho) = 1$ and $\sigma(\rho) = 2\rho(1-\rho)$ in the expressions obtained for the cumulants of $Q_T$ for general driven diffusive systems~(\ref{eq:Kappa1SM},\ref{eq:Kappa2SM},\ref{eq:Kappa3SM}), we obtain for the WASEP,
\begin{equation}
    \frac{\moy{Q_T}_c}{\sqrt{T}} \underset{T\to\infty}{\simeq}2\rb(1-\rb) \:, \qquad
     \frac{\moy{Q_T^{2}}_c}{\sqrt{T}}\underset{T\to\infty}{\simeq}\bar\rho(1-\bar\rho) \left(
      \frac{2}{\sqrt{\pi}} \e^{-(\nu \sigma'(\bar\rho))^2/4}
      + \nu \sigma'(\bar\rho) \erf \left( \frac{\nu \sigma'(\bar\rho)}{2} \right)
    \right) \:,
\end{equation}
\begin{multline}
 \frac{\moy{Q_T^{3}}_c}{\sqrt{T}}\underset{T\to\infty}{\simeq} =
  2 \nu \rb (1-\rb ) -12 \nu  \rb^2 (1-\rb)^2
  + 12 \nu \rb^2 (1-\rb)^2 \frac{\e^{-\frac{(\nu \sprb)^2}{2}}}{\pi}
  -12 \nu^3 \rb^2  (1-\rb)^2 (1-2 \rb )^2
  \\
  + 24 \nu^2 \rb^2(1-\rb )^2  (1-2\rb)  \frac{\e^{-\frac{(\nu \sprb)^2}{4}}}{\sqrt{\pi }}
  \erf\left(\frac{\nu \sprb}{2}\right)
  + 12 \nu ^3 \rb^2 (1-\rb)^2 (1-2 \rb)^2  \erf\left(\frac{\nu \sprb}{2} \right)^2
  \:.
\end{multline}
Performing the same substitution into~(\ref{eq:Kappa1tracerSM},\ref{eq:Kappa2tracerSM},\ref{eq:Kappa3tracerSM}), we obtain the cumulants of $X_T$ for the WASEP,
\begin{equation}
    \frac{\moy{X_T}_c}{\sqrt{T}} \underset{T\to\infty}{\simeq}
    2 \nu (1-\rb)
    \:, \qquad
    \frac{\moy{X_T^{2}}_c}{\sqrt{T}}\underset{T\to\infty}{\simeq}
    \frac{2 (1-\rb)}{\rb \sqrt{\pi}}
    \left(
    \e^{-\nu^2 \rb^2} +
    \nu \rb \sqrt{\pi} \: \erf (\nu \rb)
    \right)
    \:,
\end{equation}
\begin{multline}
 \frac{\moy{X_T^{3}}_c}{\sqrt{T}}\underset{T\to\infty}{\simeq} =
 \frac{2(1-\rb)}{\rb^2}
 \Bigg(
 -\frac{-3 (1-\rb) e^{-\nu ^2 \rb ^2} \left(4 \nu ^2 \rb ^2+1\right) \erf(\nu \rb)}{\sqrt{\pi}}
 \\
 +\nu   \rb (1-\rb) \left( \left(6 \nu ^2 \rb ^2+3\right) \erf(\nu  \rb )^2-6 \nu ^2 \rb^2-1\right)
 + \frac{\nu  \rb  e^{-2 \nu ^2 \rb ^2}
   \left(\pi  (3 \rb -2) e^{2 \nu ^2 \rb^2}-6 \rb +6\right)}{\pi }
  \Bigg)
  \:.
\end{multline}
In particular, inserting that $\nu = \frac{p_+ - p_-}{2} \sqrt{T}$ from the definition of the WASEP, we recover the cumulants of the ASEP~\cite{DeMasi1985SM},
\begin{equation}
   \moy{X_T} \simeq \moy{X_T^2}_c  \simeq \moy{X_T^3}_c  \simeq (p_+ - p_-)(1-\rb)T
   \:.
\end{equation}

\section{Boundary conditions for general driven diffusive systems}
\label{sec:boundary_conditions}

\subsection{Derivation from MFT}

In this section, we show that, in the case of $Q_T$, the correlation profile $\Phi$ satisfies the following boundary conditions,
\begin{equation}
\label{potential_continuity}
 \mu(\Phi)|_{0^+} - \mu(\Phi)|_{0^-}=\lambda, \quad \mu(\rho) = \int^\rho \frac{2D(r)}{\sigma(r)} \, \dd r,
\end{equation}
\begin{equation}
\label{derivative_potential_continuity}
\partial_x \mu(\Phi)|_{0^+} = \partial_x \mu(\Phi)|_{0^-}
\:,
\end{equation}
for any driven diffusive system. We adapt the procedure used in Ref.~\cite{Grabsch:2024SM} for non-driven systems, and introduce
\begin{equation}
    q(x, t) = \hat{q}(x, 1 - t), \quad
    p(x, t) = -\hat{p}(x, 1 - t) + \mu(\hat{q}(x, 1 - t))-2\nu x
    \:.
\end{equation}
Inserting these relations into the MFT equations~(\ref{eq:MFT_q},\ref{eq:MFT_p}), we find that $\hat{q}$ and $\hat{p}$ obey the exact same equations,
\begin{align}
    \partial_t  \hat{q}
    &= \partial_x \left[
      D( \hat{q}) \partial_x  \hat{q} - \sigma( \hat{q}) \partial_x  \hat{p} - \nu \sigma( \hat{q})
      \right]
      \:,
    \\
    \partial_t  \hat{p}
    &= - D( \hat{q})\partial_x^2  \hat{p} - \frac{1}{2} \sigma'( \hat{q}) (\partial_x  \hat{p})^2
      - \nu \sigma'( \hat{q}) \partial_x  \hat{p}
      \:,
\end{align}
but with the new initial and final condition deduced from~\eqref{eq:InitTimeSM},
\begin{equation}
  \label{eq:MFTboundtimereversal}
   \hat{p}(x,0) = -\lambda \Theta(x) + \mu(\hat{q}(x,0))-2\nu x
  \:,
  \quad
   \hat{p}(x,1) = -\lambda \Theta(x) + \mu(\rb)-2\nu x
  \:.
\end{equation}
The key point is that $\hat{p}$ obeys an anti-diffusion equation, so that it is smooth at $t=0$, hence,
\begin{equation}
    \hat{p}(0^{-},0) = \hat{p}(0^{+},0), \qquad \partial_x\hat{p}(0^{-},0) = \partial_x\hat{p}(0^{+},0)
    \:.
\end{equation}
Using the expression of $\hat{p}(x,0)$~\eqref{eq:MFTboundtimereversal} and using that $\hat{q}(x,0) = q(x,1) = \Phi(x)$, we obtain the boundary conditions~(\ref{potential_continuity},\ref{derivative_potential_continuity}) announced above. Note that in the case of the WASEP, $D(\rho) = 1$ and $\sigma(\rho) = 2 \rho(1-\rho)$, these equations reduce to~\eqref{eq:lambda_wasep} obtained from microscopic considerations.

\subsection{A conjecture for the cumulant generating function}
\label{sec:Conjec}

In the case of the WASEP discussed in Section~\ref{eq:WASEPmacro}, the two boundary conditions~(\ref{potential_continuity},\ref{derivative_potential_continuity}) are completed by a third condition which relates $\hat\psi$ to $\Phi(0^\pm)$ and $\partial_x \Phi(0^\pm)$. In practice, this last relation greatly simplifies the computation of the cumulants from the correlation profile $\Phi$ since it bypasses the computation of the spatial integral in~\eqref{eq:dpsidlambda}. The WASEP relation~\eqref{eq:psi_wasep} can actually be written in terms of $D$ and $\sigma$ as
\begin{equation}
    \label{eq:psi_continuity_driven}
    \hat{\psi}+\nu\partial_\nu\hat{\psi} = -2\left(\partial_x \mu(\Phi)|_{0^+}-2\nu\right)\int^{\Phi(0^+)}_{\Phi(0^-)}D(r)\dd r
    \:.
\end{equation}
We conjecture that this relation is valid for any driven diffusive system. This conjecture is verified for the WASEP by construction, and for any driven diffusive system up to order $3$ in $\lambda$ from our results on the cumulants~(\ref{eq:Kappa1SM},\ref{eq:Kappa2SM},\ref{eq:Kappa3SM}) and the correlation profiles~(\ref{q1T1},\ref{q2T1}). In the non-driven case $\nu=0$, it reduces to the expression written in~\cite{Grabsch:2024SM}.

\bigskip

Remarkably, the relation~\eqref{eq:psi_continuity_driven} is invariant under the duality mapping described in Section~\ref{subsection:duality}, as we now proceed to show. The profile $\tilde{\Phi}$ in the dual system is related to the profile $\Phi$ in the original system by
\begin{equation}
    \tilde{\Phi}(y(x)) = \frac{1}{\Phi(x)} \, \quad y(x,t) = \int_0^x \Phi(x') \dd x' \:,
\end{equation}
and the coefficients $D(\rho)$, $\sigma(\rho)$ and $\nu$ are mapped according to~\eqref{eq:MappingTrCoefsSM}. Applying that mapping to the term at the right hand side of the conjectured relation~\eqref{eq:psi_continuity_driven}, we obtain for the chemical potential
\begin{equation}
    \partial_x \mu(q)= -\frac{\partial_y\tilde{q}}{\tilde{q}^{3}} \mu' \left(\frac{1}{\tilde{q}} \right) = \frac{\partial_y\tilde{q}}{\tilde{q}^{3}}\frac{2D(\frac{1}{\tilde{q}})}{\sigma(\frac{1}{\tilde{q}})}=-\partial_y \tilde{\mu}(\tilde{q}),
\end{equation}
where $\tilde{\mu}$ is the chemical potential of the dual model.  Setting the change of variable $u=\frac{1}{r}$, in the integral,
\begin{equation}
    \int^{\Phi(0^+)}_{\Phi(0^-)}D(r)\dd r = -\int^{\frac{1}{\Phi(0^+)}}_{\frac{1}{\Phi(0^-)}}D \left(\frac{1}{u} \right)
    \frac{\dd u}{u^2}=  -\int^{\tilde \Phi(0^+)}_{\tilde \Phi(0^-)}\Dt(u)\dd u
    \:.
\end{equation}
All together, we deduce that the term on the r.h.s. of \eqref{eq:psi_continuity_driven} is invariant under the duality mapping, which is consistent with the fact that $\hat{\psi}$ is also invariant. This further supports the validity of the relation~\eqref{eq:psi_continuity_driven}.

\bibliographystyle{apsrev4-1}

\end{document}